\documentclass[aps,prl,letterpaper,10pt,twocolumn,notitlepage,superscriptaddress]{revtex4-1}
\usepackage[colorlinks=true,allcolors=blue!50!black]{hyperref}
\usepackage{mathptmx}
\usepackage{amssymb}
\usepackage{amsmath}
\usepackage{graphicx}
\usepackage{subfigure}
\usepackage{amsfonts}
\usepackage{xcolor}
\usepackage{epsfig}
\usepackage{color}
\usepackage{bbm}
\usepackage{bm}
\usepackage{tabularx}
\usepackage{multirow}
\usepackage{mathtools}
\usepackage{xfrac}
\DeclareMathAlphabet{\mathcal}{OMS}{cmsy}{m}{n}
\newcommand{\ket}[1]{\vert{#1}\rangle} 
\newcommand{\bra}[1]{\langle{#1}\vert} 
 
\newcommand{\trans}[2]{\ket{#1}\!\bra{#2}}
\newcommand{\mean}[1]{\langle #1 \rangle}

\newcommand{\abs}[1]{\left|#1\right|}
\newcommand{\sighat}{\hat{\sigma}}

\newcommand{\xzpm}{z_{\rm zp}}

\newcommand{\half}{\frac{1}{2}}

\newcommand{\tr}[1]{\mathrm{Tr}#1}
\newcommand{\hatd}[1]{\hat{#1}^{\dagger}}

\begin{document}
\title{Quantum Effects in a Mechanically Modulated Single Photon Emitter}

\author{Mehdi Abdi}
\email{mehabdi@gmail.com}
\affiliation{Department of Physics, Isfahan University of Technology, Isfahan 84156-83111, Iran}
\affiliation{Institute of Theoretical Physics and IQST, Albert-Einstein-Allee 11, Ulm University, 89069 Ulm, Germany}

\author{Martin B. Plenio}
\affiliation{Institute of Theoretical Physics and IQST, Albert-Einstein-Allee 11, Ulm University, 89069 Ulm, Germany}

\begin{abstract}
Recent observation of quantum emitters in monolayers of hexagonal boron nitride (h-BN) has provided a novel platform for optomechanical experiments where the single-photon emitters can couple to the motion of freely suspended h-BN membrane.
Here, we propose a scheme where the electronic degree of freedom of an embedded color center is coupled to the motion of the hosting h-BN resonator via dispersive forces.
We show that the coupling of membrane vibrations to the electronic degree of freedom of the emitter can reach the strong regime.
By suitable driving of a three-level $\Lambda$-system composed of two spin degrees of freedom in the electronic ground state as well as an isolated excited state of the emitter a multiple electromagnetically induced transparency spectrum becomes available.
The experimental feasibility of the efficient vibrational ground-state cooling of the membrane via quantum interference effects in the two-color drive scheme is numerically confirmed.
More interestingly, the emission spectrum of the defect exhibits a frequency comb with frequency spacings as small as the fundamental vibrational mode, which finds applications in high-precision spectroscopy.
\end{abstract}

\maketitle

%
%
\textit{Introduction.---}%
Quantum metrology relies on the resources provided in quantum systems to beat the standard limits set by quantum physics~\cite{Giovannetti2011}. High-precision spectroscopy that has become available thanks to the development of optical frequency combs~\cite{Holzwarth2000, DelHaye2007} is a powerful tool which may be improved further by employing quantum resources~\cite{Pinel2012}.
The optical quantum communication as another major quantum technology relies on efficient and broadband quantum memories where the photons can be faithfully stored and retrieved~\cite{Lvovsky2009, Reiserer2015}.
The discovery of single-photon emitters (SPE) in two dimensional (2D) systems such as WSe$_2$~\cite{Srivastava2015} and h-BN~\cite{Tran2016a} represents a significant step forward for quantum technology where miniaturization of the devices without compromising their performance is of crucial importance. Such systems also hold considerable promise in the field of optomechanics and the related technologies~\cite{Abdi2017, Aharonovich2017}.
The coupling of SPEs to mechanical vibrations has already proven successful with outstanding achievements by diamond color centers~\cite{Wilson-Rae2004, Kepesidis2013, Ovartchaiyapong2014, Barfuss2015, Lemonde2018}.
Nonetheless, the extremely low mass and high quality mechanical resonators achievable in free-standing 2D membranes make them promising objects for sensing weak forces and small displacements~\cite{Moser2014, Singh2014, Muschik2014, Castellanos-Gomez2015, Reserbat2016, Morell2016, Li2016, Falin2017, Zheng2017}. These are capabilities that allow for the exploration of various problems in fundamental physics, e.g. the validation of the exotic decoherence models~\cite{Bassi2013, Riedinger2016, Abdi2016} and are advantageous in technological applications~\cite{Weis2010, Safavi2011}.

Here, we propose an electromechanical system where an emitter embedded in a free-standing hexagonal boron nitride (h-BN) membrane couples to its vibrational modes via the vacuum dispersive forces.
The setup even reaches a regime where the strength of the coupling between electronic states of the emitter and mechanical oscillations of the h-BN membrane exceeds their dissipation rates and characteristic frequencies, the so-called ultrastrong coupling regime.
By realizing a $\Lambda$-system in the emitter and employing a two-color drive scheme we study various aspects in the dynamics of our system. The scheme is tuned to the electromagnetically induced transparency (EIT) regime, where the quantum interference between two different paths opens-up a transparency window in the single-atom level~\cite{Mucke2010, Kampschulte2010, Slodicka2010}. Due to the ultrastrong coupling to the mechanical mode the EIT signal exhibits multiple absorption and transparency windows with distances matching the mechanical frequency.
The emission spectrum of the emitter in this working regime---when thermal occupation of the mechanical mode are made sufficiently small---is a tightly spaced frequency comb. The high emission rate of photons~\cite{Tran2016a} allows for formation of a time-bin entangled string of photons that spans the comb, by employing a sequential entangling protocol~\cite{Schon2005, Schon2007}. This highly entangled configuration has many application in quantum metrology such as quantum boosted optical spectroscopy~\cite{Leibfried2004, Obrien2009, Reimer2016}.
Furthermore, we propose to achieve the ground state mechanical cooling via the same emitter and by applying the EIT cooling technique originally developed for trapped ions~\cite{Morigi2000}. When the optical drives are both far off-resonance and blue detuned, the absorption spectrum assumes a Fano-like shape, hence, allowing a resolved-sideband transition via the emitter which then cools the mechanical vibrations down to the ground state. By merging this cooling process with the EIT signal in a pulsed scheme we show that quantum effects are robust against mechanical thermal noise.

\begin{figure}[tb]
\includegraphics[width=\columnwidth]{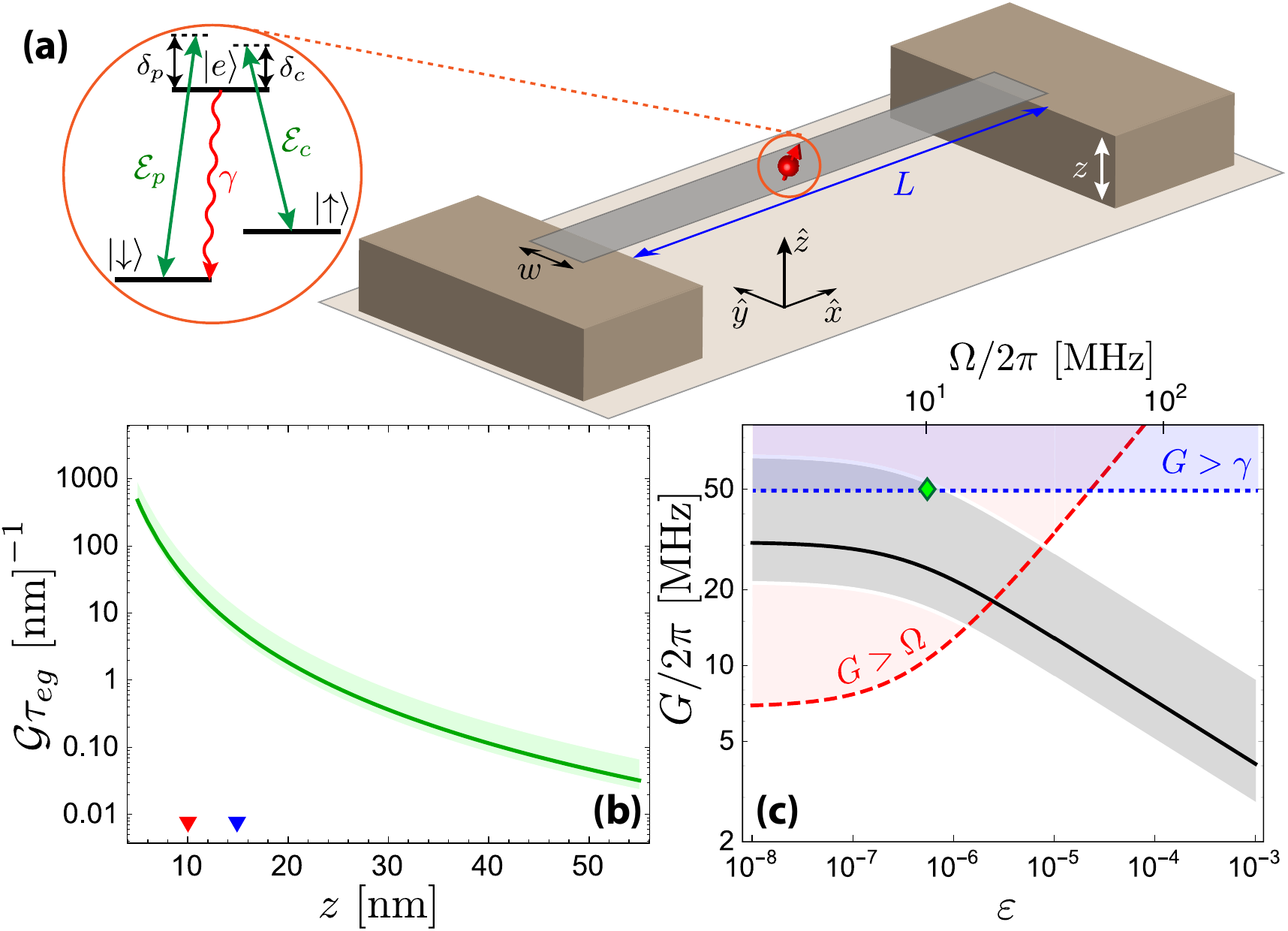}
\caption{%
(a) Sketch for the proposed setup: The h-BN nanoribbon hosting a color center whose electronic ground and excited states form a $\Lambda$ system.
(b) Normalized dispersion force frequency pull of an h-BN emitter as a function of its distance from the surface; the solid green line corresponds to the average emission energies $1.95$~eV while the shade depicts the inhomogenous broadening of the emitters. The blue and red triangles show the distances attained experimentally~\cite{Reserbat2016} and assumed in this paper, respectively.
(c) Absolute electromechanical coupling rate for a monolayer h-BN strip versus clamping strain for the parameter values $z=10$~nm, $L=1~\mu$m, $w=10$~nm, and $\tau_{eg} = 3.2$~ns. For reference, top axis gives the corresponding mechanical frequency.
The dashed red and dotted blue lines separate the ultrastrong coupling region.
The green diamond indicates our working point in the rest of paper. 
}%
\label{fig:scheme}%
\end{figure}

%
%
\textit{Model.---}%
A sketch of our proposed setup is given in Fig.~\ref{fig:scheme}(a). The h-BN monolayer membrane is suspended at a distance $z$ from the substrate surface. We single out three states of the emitter that realize a $\Lambda$-system composed of two electronic ground states (corresponding to two different spin states with energy splitting $\Delta_0$) and a bright excited state~\cite{Abdi2018}.
The reflection of virtual photons from the underlying surface induces off-resonant transitions in the emitter. This will modify transition energy of the emitter, and consequently, induce a dispersive force on it~\cite{Buhmann2004}.
When the emitter is hosted by an oscillating object the lowest emitter transition frequency $\omega_{eg}$ will depend on its position. For emitters sufficiently close to the surface of the substrate ($z\ll\lambda$), in the first order of approximation, the change in the transition frequency is given by (see supplemental information for a derivation)
\begin{equation}
\delta\omega_{eg}(z) \approx \frac{3}{32}\frac{c^3}{\tau_{eg}\omega_{eg}^3z^3}\frac{\abs{\epsilon(\omega_{eg})}^2-1}{\abs{\epsilon(\omega_{eg})+1}^2},
\end{equation}
where $\omega_{eg}$ and $\tau_{eg} = 1/\gamma$ are the \textit{free-space} transition frequency and the excited state lifetime, respectively. Here, $\gamma$ is the photon emission rate and $\epsilon(\omega)$ is the electric permittivity of the substrate material.
The dependence of frequency shift on the displacement results-in the dispersion-force-induced frequency shift $\mathcal{G} = \partial_z\delta\omega_{eg}$, which provides the coupling between membrane vibrations and electronic state of the emitter. In Fig.~\ref{fig:scheme}(b) the normalized value of $\mathcal{G}$ versus distance from the surface is given considering the inhomogeneous broadening of the observed emission frequencies $\omega_{eg}$.

%
%
\textit{Parameters.---}%
Apart from distance to the substrate, the absolute value of the electromechanical coupling rate $G \equiv \mathcal{G}\xzpm$ depends on the emission frequency $\omega_{eg}$, emission rate $\gamma$, and amplitude of the zero-point fluctuations $\xzpm = \sqrt{\hbar/2m\Omega}$.
We consider a long rectangular strip of dimensions $L\cdot w\cdot h$ clamped at both ends with tensile strain $\varepsilon$ [Fig.~\ref{fig:scheme}]. The normal modes of this geometry are found by solving the elastic wave equation with proper boundary conditions~\cite{suppinfo, Landau1975}.
For a monolayer hBN nano-ribbon with determined dimensions the coupling rate can be tuned by varying the clamping strain.
The coupling strength for the fundamental vibrational mode of a $1~\mu$m ribbon to the electronic degree of freedom of an emitter about the center of the strip is plotted in Fig.~\ref{fig:scheme}(c). We notice that coupling rates as high as $\simeq 50$~MHz are realistic in this setup. In our calculations we have assumed a strip of width $w=10$~nm, a free-space emission lifetime of $\tau_{eg} \approx 3.2$~ns for the emitter~\cite{Jungwirth2016}, and $z=10$~nm, which is well-within reach~\cite{Reserbat2016}. With these parameters the system achieves the strong coupling regime, i.e. $\abs{G} \gtrsim 1/\tau_{eg}$.
In the remainder of the paper we discuss the manifestations of such strong coupling in the optical properties of the emitter.

%
%
\textit{Cooling.---}%
To observe the mechanical manifestations in the quantum state of the emitted photons the prohibiting mechanical thermal noise, which is the dominant destructive effect, must be overcome first.
One thus has to work at low temperatures
In our setup, however, the thermal occupation number of the vibrational fundamental mode is still considerable even at temperatures as low as $T=0.1$~K.
Hence, a cooling mechanism must be invoked.
At the chosen working point that provides us the interesting effects discussed below, the conditions for ground state cooling by the sideband cooling method are not satisfied~\cite{Rabl2010}. Most importantly the mechanical sidebands are not resolved $\tau_{eg} \Omega > 1$ in the spectrum of the qubit. Yet, the emitter is strongly coupled to the mechanical mode and the back-actions are not suppressed~\cite{Degenfeld2016}.
We instead employ an EIT cooling technique---originally developed for trapped ions~\cite{Morigi2000}---and show that this method efficiently cools the mechanical mode down to its ground-state. 

\begin{figure}[b]
\includegraphics[width=\columnwidth]{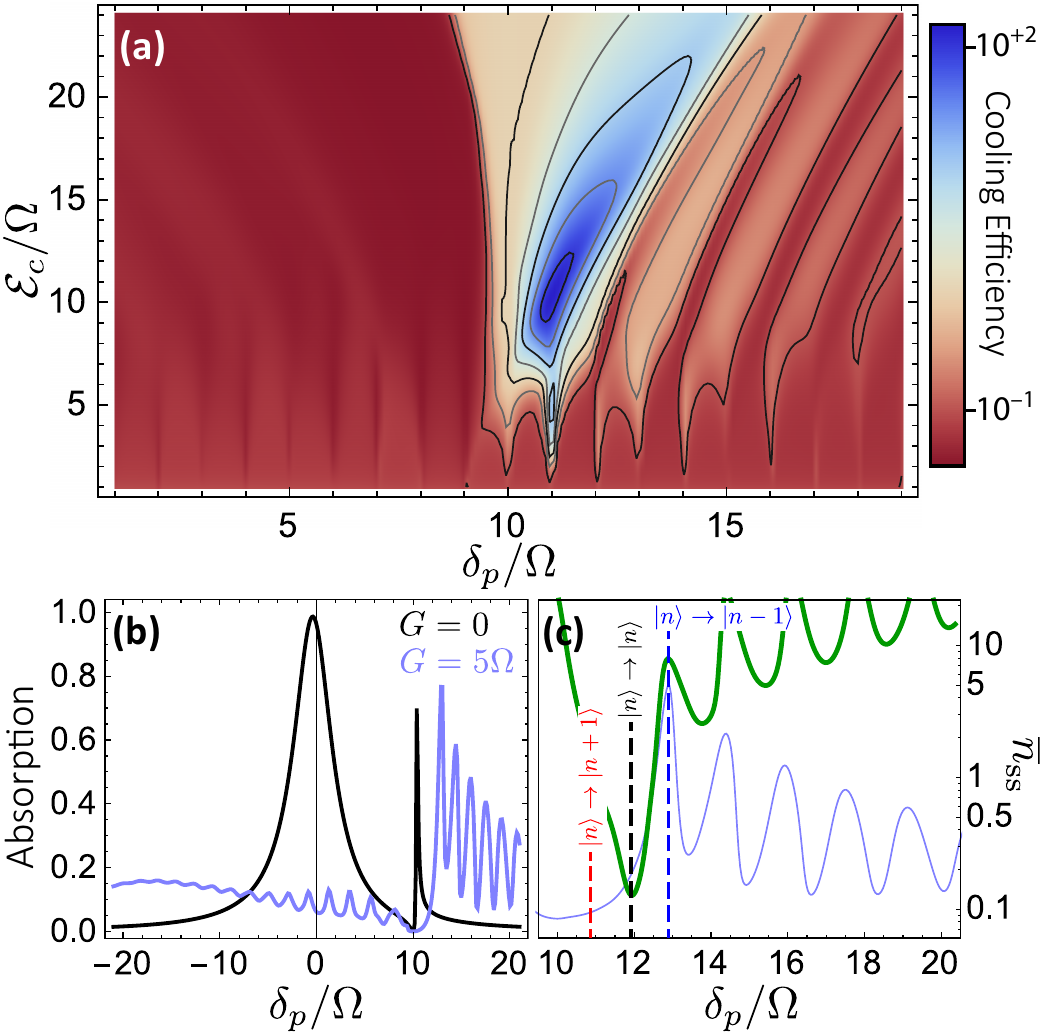}
\caption{%
(a) EIT cooling efficiency as function of probe detuning and the control Rabi frequency. The probe Rabi frequency is set at $\mathcal{E}_p=0.1\mathcal{E}_c$ and $\delta_c = +10\Omega$.
(b) Absorption spectrum of the probe field in the absence (black) and presence (blue) of coupling to the mechanical mode.
(c) Variations of the steady-state mechanical occupation number with probe drive detuning (solid green line). The absorption spectrum as well as the phonon assisted transitions are also shown for reference (see the text for a discussion). In (b) and (c) $\mathcal{E}_c = 14\Omega$ and $\bar{N}_\Omega = 210$ corresponding to the temperature $T=0.1$~K.
}%
\label{fig:cooling}%
\end{figure}

The $\Lambda$-system is driven by one optical control drive tuned at the $\trans{\uparrow}{e}$ transition, while another optical probe drives the $\trans{\downarrow}{e}$ transition [Fig.~\ref{fig:scheme}(a)]. Hamiltonian of this system in the rotating frame of the two transition frequencies~\cite{Rebic2004} (with adopting the notation $\sighat_{ij}\equiv\trans{i}{j}$) is given by
\begin{align}
\hat H_\Lambda = &-\delta_p\sighat_{ee} -(\delta_p -\delta_c)\sighat_{\uparrow\uparrow} +\Omega\hatd b\hat b +G\sighat_{ee}(\hat b+\hatd b) \nonumber\\
&+\mathcal{E}_p\sighat_{e\downarrow} +\mathcal{E}_c\sighat_{e\uparrow} +\mathrm{H.c.},
\label{lambda}
\end{align}
where $\delta_p \equiv \omega_p - \omega_{eg}$ and $\delta_c \equiv \omega_c - \omega_{eg} +\Delta_0$ are detuning of the probe and control fields from their respective transitions.
The second line gives the drive Hamiltonian with the probe and control Rabi frequencies $\mathcal{E}_c$ and $\mathcal{E}_p$ such that $\mathcal{E}_c \gg \mathcal{E}_p$.
For EIT cooling the control transition of the $\Lambda$-system $\sighat_{\uparrow e}$ is driven blue detuned from the resonance with a proper Rabi frequency such that the large decay rate of the excited state  is significantly reduced by quantum interference effects in a Fano-like resonance spectrum for the probe field. This tunes the system into the regime where the phonon annihilation processes are significantly more probable than the phonon creation leading to efficient mechanical cooling [Fig.~\ref{fig:cooling}].

The feasibility of ground-state cooling of an h-BN nano-ribbon via dispersive electromechanical coupling is examined numerically by finding the steady-state solution of the Lindblad master equation $\dot\rho = \mathcal{L}\rho$ with the Liouvillian
\begin{align}
\mathcal{L}\rho &=\! -i[\hat{H}_\Lambda,\rho] +\frac{\gamma}{2}\mathcal{D}_{\sighat_{\downarrow e}}\rho 
+\frac{\Gamma}{2}\Big[(\bar{N}_{\Omega}+1)\mathcal{D}_{\hat b}\hspace{0.5mm}\rho +\!\bar{N}_{\Omega}\mathcal{D}_{\hat b^\dag}\rho\Big], \nonumber\\
\mathcal{D}_{\hat{o}}\rho &= 2\hat{o}\rho\hat{o}^\dag -\hat{o}^\dag\hat{o}\rho -\rho\hat{o}^\dag\hat{o},
\label{liouv}
\end{align}
where $\Gamma \equiv \Omega/Q$ is the damping rate of the mechanical mode with quality factor $Q$ and $\bar{N}_\omega = \big(\exp\{\frac{\hbar\omega}{k_{\rm B}T}\}-1\big)^{-1}$ is the thermal occupation number of a bosonic mode with frequency $\omega$ at temperature $T$. ($k_{\rm B}$ is the Boltzmann constant).
Throughout this work we take a mechanical quality factor of $Q=7000$ which is within the reach~\cite{Cartamil-Bueno2017, Shandilya2018}.
In Fig.~\ref{fig:cooling} the numerical results are summarized where we present the final occupation number of the mechanical mode $\overline{n}_{\rm ss} \equiv \tr\{\hatd b\hat b \rho_{\rm ss}\}$ and the cooling efficiency $\bar{N}_\Omega/\overline{n}_{\rm ss}$.
$\rho_{\rm ss}$ is the steady-state density matrix ($\mathcal{L}\rho_{\rm ss}=0$). The complicated cooling pattern in Fig.~\ref{fig:cooling}(a) is understood by generalizing the EIT cooling mechanism which results from the narrowed absorption spectrum of the $\Lambda$-system at large detunings to the multiple absorption pattern resulting from the strong coupling to the mechanical mode [Fig.~\ref{fig:cooling}(b)]. This is clearly observable in Fig.~\ref{fig:cooling}(c) where we plot $\overline{n}_{\rm ss}$ alongside the system absorption.
When the probe field is driven at the proper detuning values the phonon absorption processes are enhanced while the phonon preserving and creating transitions are largely suppressed. The minimum steady-state phonon number ($\min[\overline{n}_{\rm ss}] \approx 0.12$) is obtained when the probe is driven next to the first Fano peak, the global maximum of the spectrum.
It is noteworthy that the timescale required for achieving the ground-state cooled mechanical mode, $t_{\rm ss}$, is quite short. The time needed for the system to reach its steady-state is determined by its slowest decay rate that, in turn, is given by the Liouvillian eigenvalue with smallest nonzero real part. This quantity is numerically evaluated and we find that $t_{\rm ss} \approx 7\tau_{\rm m}$ where $\tau_{\rm m}\equiv 2\pi/\Omega$ is the mechanical period.
Such a fast cooling scheme is beneficial in avoiding prohibitive effects such as laser heating~\cite{Meenehan2015, suppinfo}.

\begin{figure}[tb]
\includegraphics[width=\columnwidth]{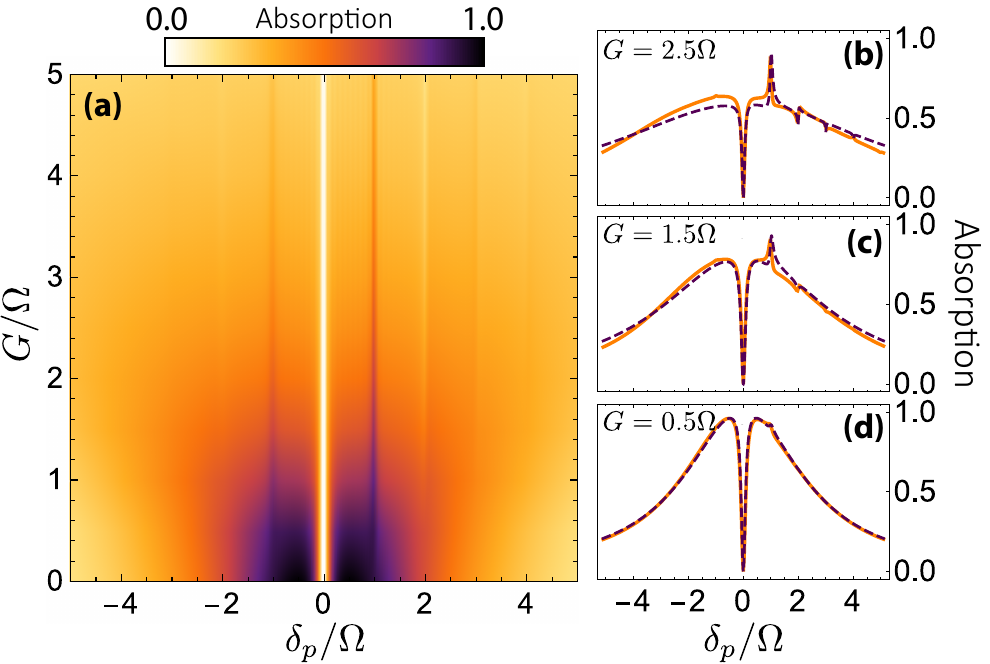}
\caption{%
Electromagnetically induced signal:
(a) At $t=10\tau_{\rm m}$ as a function of probe detuning and the strength of coupling to the mechanical mode ($\bar N_\Omega = 210$).
In (b)-(d) the ideal ($\bar N_\Omega = 0$) steady-state EIT signal is illustrated at three different coupling rates: The solid lines are for numerical computation, while the dashed lines indicate analytical expression given in Eq.~(\ref{EIT}). The parameters used in these plots are $\mathcal{E}_c = 10\mathcal{E}_p = \Omega$ and $\delta_c=0$. 
}%
\label{fig:EIT}%
\end{figure}

%
%
\textit{Multiple EIT.---}%
We now consider the electromagnetically induced transparency scheme. The EIT signal in our setup assumes multiple dips and peaks~\cite{Wang2014, Alotaibi2014} that root back to an interesting property of the setup as will become clear shortly.
The standard EIT signal which is proportional to absorption spectrum of $\sighat_{\downarrow e}$ manifests itself when the control field is strongly driving on-resonance ($\delta_c = 0$, $\mathcal{E}_c \gg \mathcal{E}_p$).
However, the thermal mechanical noise washes out the signal. The mechanical mode thus must be cooled down close to its ground state as studied in the preceding section.
This can be provided in a continuous wave (CW) fashion by a second emitter~\cite{Abdi2017}. Alternatively, a pulsed scheme can be used where the emitter is first employed for cooling then it is driven in the EIT regime for timescales shorter than the mechanical thermalization.
The system behavior in the second scenario is studied numerically by solving the time dependent master equation with Liouvillian (\ref{liouv}).
The results are presented in Fig.~\ref{fig:EIT}(a) where the modification of the absorption spectrum of $\trans{\downarrow}{e}$ with the coupling strength is presented in a density plot. The plot indicates that alongside the normal transparency window occurring at $\delta_p=0$ the system exhibits multiple valleys and peaks at mechanical sidebands $\delta_p = k\Omega~~(k=\pm 1, \pm 2, \cdots)$ when the emitter is strongly interacting with the mechanical mode. The stronger the coupling, the more sidebands appear.

To better understand the behavior of the EIT signal, we derive an approximate analytical expression for $\bra{e}\rho_{\rm ss}\ket\downarrow \equiv \rho_{\downarrow e}$ which is proportional to the probe absorption spectrum.
We first apply the unitary transformation $\hat U = \exp\{\frac{G}{\Omega}\sighat_{ee}(\hatd b -\hat b)\}$ to the system Hamiltonian in (\ref{lambda}). Then by assuming no back-action from the emitter on the mechanical mode we calculate the steady-state mechanical correlation functions. This is justified for small drives $\mathcal{E}_p, \mathcal{E}_c < \gamma$ for which the excited state population remains negligible.
This brings us to the approximate relation~\cite{suppinfo}:
\begin{align}
\rho_{\downarrow e} \approx \frac{i\mathcal{E}_p\alpha}{\gamma/2 -i\tilde\delta_p +\mathcal{E}_c^2\alpha^2 \sum_{n=0}^\infty\sum_{k=0}^n \binom{n}{k}\frac{L(n,k)}{2n!}},
\label{EIT}
\end{align}
where $\alpha \equiv \exp\{-\half(\frac{G}{\Omega})^2(2\bar{N}_{\Omega}+1)\}$ is the signal attenuation coefficient as a consequence of the mechanical vibrations and shows that for $\bar N_\Omega \gg 1$ the signal will be completely demolished. Also we have introduced the Lorentzian function $L(n,k)\equiv (\frac{G}{\Omega})^{2n}\frac{\bar{N}_{\Omega}^{n-k}(\bar{N}_{\Omega}+1)^k}{n\Gamma/2 -i[\delta_p+(n-2k)\Omega]}$ illuminating the mechanical sidebands. We emphasize that Eq.~(\ref{EIT}) is only valid for small coupling rates $G \lesssim \Omega$ where the dynamics of the mechanical mode is sufficiently independent of the emitter's excited state.

In Figs.~\ref{fig:EIT}(b-d) the analytical expression is compared to the steady-state numerical absorption spectrums. Although the analytical expression (\ref{EIT}) captures most features of the spectrum, one obviously finds the build-up of discrepancies as the coupling rate increases. This includes the small blue sideband ($\delta_p = -\Omega$) spectral feature that is an indication of the mechanical back-action.
Furthermore, the setup exhibits multiple sharp dispersive features in the imaginary part of the $\sighat_{\downarrow e}$ transition located at integer multiples of the mechanical frequency, see the supplemental material~\cite{suppinfo}.
In practice, one needs to enhance the signal by either placing the setup in a cavity~\cite{Mucke2010} or focusing a light beam on the emitter~\cite{Slodicka2010}. The former can alternatively be realized in an h-BN photonic crystal cavity~\cite{Kim2018}.

\begin{figure}[b]
\includegraphics[width=\columnwidth]{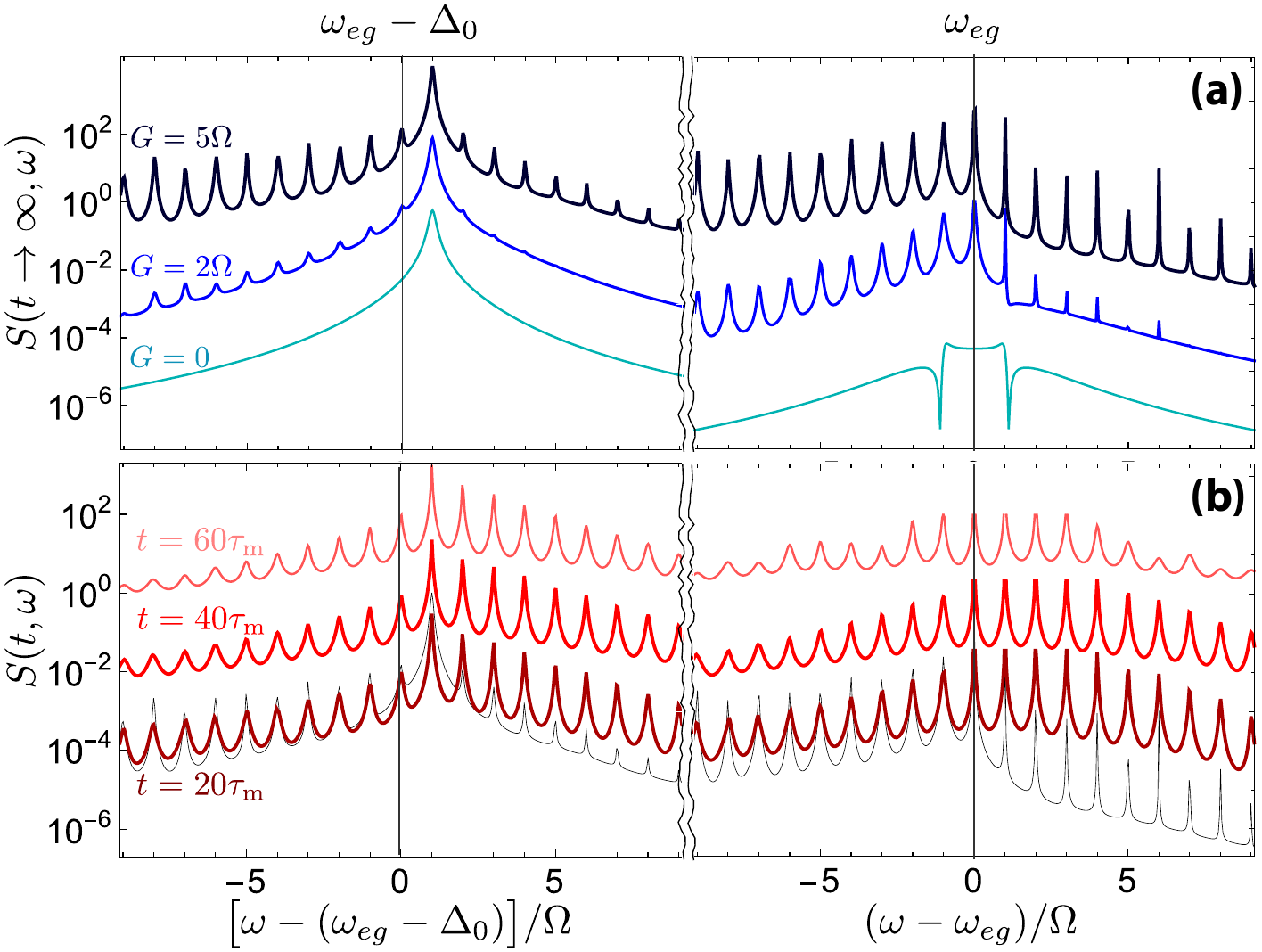}
\caption{%
(a) Steady-state RFS of the emitter $S(t\rightarrow\infty,\omega)$ at the two resonance frequencies $\omega_{eg}$ and $\omega_{eg}-\Delta_0$ for three different coupling rates ($\bar N_\Omega=0$).
(b) Time-resolved RFS for $G=5\Omega$ and $\bar N_\Omega=210$ at different times with system initialized with the ground-state. The ideal steady-state spectrum is also shown for comparison (the  thin line).
In these plots we have set $\delta_c = 0$, $\delta_p = \Omega$, and $\mathcal{E}_c = 10\mathcal{E}_p = \Omega$.
The curves are shifted vertically for clarity.
}%
\label{fig:spectra}%
\end{figure}

%
%
\textit{Frequency Comb.---}%
We next study the resonance fluoresce spectrum (RFS) of the emitter. The spectrum about emitter transition frequencies is given by
\begin{align}
S(t,\omega) = \int_{\!-\infty}^{\infty}\!&d\tau e^{-i\omega\tau} \Big[e^{i\omega_{eg}\tau}\mean{\sighat_{e\downarrow}(t+\tau)\sighat_{\downarrow e}(t)} \nonumber\\
&+e^{i(\omega_{eg}-\Delta_0)\tau}\mean{\sighat_{e\uparrow}(t+\tau)\sighat_{\uparrow e}(t)}\Big],
\end{align}
where the two-time correlation functions are calculated by the quantum regression theorem~\cite{Narducci1990, Carmichael1999}.
We first numerically evaluate the steady-state of this function for the case of $\bar N_\Omega = 0$ at the ultrastrong $G \gtrsim \Omega$ coupling regime. This can be realized by employing a second emitter on the membrane for the purpose of cooling.  We find that in this regime, the mechanical mode manifests itself as sideband peaks in the emission spectrum, first at the redshifted frequencies and then in the blueshift side of the resonances as the coupling rate is raised. Eventually, for sufficiently large coupling strength a frequency comb with frequency distances determined by the mechanical mode frequency $\Omega$ emerges [Fig.~\ref{fig:spectra}(a)].
One alternatively adopts the pulsed scenario. In fact, the time for formation of the comb is given by the coupling strength and emitter decay rate. Hence, a cooling pulse with duration of a few $\tau_{\rm m}$ followed by an EIT pulse as short as a few mechanical period will arrange the frequency comb. For long waiting times thermalization of the mechanical mode will broaden the comb teeth and eventually flatten them one by one starting from the farther sidebands. This is clear from Fig.~\ref{fig:spectra}(b) where we plot the RFS for three different pulse durations.
As the system evolves the mechanical occupation number, $\overline{n}(t)\equiv \tr\{\hatd b\hat b\rho(t)\}$, rises  from the initial value $0.12$ to $2.78$, $5.42$, and $7.98$ for 20, 40, and 60 mechanical periods, respectively.
Such equally spaced and tightly arranged frequency combs are useful in precision measurements and metrology as well as light storage quantum memory~\cite{Afzelius2010}.

The comb arising from the strongly mixed electronic--mechanical system provides a quantum memory based on photon echo~\cite{Afzelius2010, Moiseev2010, Sabooni2013}. The ratio of peak separation and linewidth, the comb finesse $\mathcal{F}$, is a figure of merit in such schemes. This quantity can assume very large values in our setup both in the steady-state scenario ($\mathcal{F}=Q\sim 10^4$) and short EIT pulses (e.g. $\mathcal{F}= Q[\overline{n}(10\tau_{\rm m})+1]^{-1}\sim10^3$).
A high efficiency quantum memory, however, requires optimal engineering of the comb structure and suppression of the spectral dispersion~\cite{Moiseev2012, Arsalanov2017}. This, in principle, can be done by employing a higher vibrational mode of the membrane that couples to the emitter, and hence, produces two absorbing lines on the sides of the frequency comb. Such spectral sidebands compensate for the dispersion of the signal photons by providing slow-light effects, and thus, enhance the efficiency of the memory~\cite{Moiseev2012}.
Another interesting feature of the comb is the possibility of creating a time-frequency chain of entangled photons. Since the single photons emitted by the color center assume any of the peaked frequencies depicted in Fig.~\ref{fig:spectra} with a finite probability, application of a photon entangling protocol that employs the same $\Lambda$-system as an ancillary will create a comb of time-bin entangled photons~\cite{Schon2007, Rao2015, Dhand2018}. These are very useful in super-precision quantum imaging and metrology~\cite{Giovannetti2011, Chen2014}.
\begin{acknowledgments}
This work was supported by the ERC Synergy grant BioQ and the EU project AsteriQs.
The authors also acknowledge support
by the state of Baden-W{\"u}rttemberg through bwHPC.
\end{acknowledgments}

%
%
\bibliography{dispersion}

\begin{thebibliography}{65}%
\makeatletter
\providecommand \@ifxundefined [1]{%
 \@ifx{#1\undefined}
}%
\providecommand \@ifnum [1]{%
 \ifnum #1\expandafter \@firstoftwo
 \else \expandafter \@secondoftwo
 \fi
}%
\providecommand \@ifx [1]{%
 \ifx #1\expandafter \@firstoftwo
 \else \expandafter \@secondoftwo
 \fi
}%
\providecommand \natexlab [1]{#1}%
\providecommand \enquote  [1]{``#1''}%
\providecommand \bibnamefont  [1]{#1}%
\providecommand \bibfnamefont [1]{#1}%
\providecommand \citenamefont [1]{#1}%
\providecommand \href@noop [0]{\@secondoftwo}%
\providecommand \href [0]{\begingroup \@sanitize@url \@href}%
\providecommand \@href[1]{\@@startlink{#1}\@@href}%
\providecommand \@@href[1]{\endgroup#1\@@endlink}%
\providecommand \@sanitize@url [0]{\catcode `\\12\catcode `\$12\catcode
  `\&12\catcode `\#12\catcode `\^12\catcode `\_12\catcode `\%12\relax}%
\providecommand \@@startlink[1]{}%
\providecommand \@@endlink[0]{}%
\providecommand \url  [0]{\begingroup\@sanitize@url \@url }%
\providecommand \@url [1]{\endgroup\@href {#1}{\urlprefix }}%
\providecommand \urlprefix  [0]{URL }%
\providecommand \Eprint [0]{\href }%
\providecommand \doibase [0]{http://dx.doi.org/}%
\providecommand \selectlanguage [0]{\@gobble}%
\providecommand \bibinfo  [0]{\@secondoftwo}%
\providecommand \bibfield  [0]{\@secondoftwo}%
\providecommand \translation [1]{[#1]}%
\providecommand \BibitemOpen [0]{}%
\providecommand \bibitemStop [0]{}%
\providecommand \bibitemNoStop [0]{.\EOS\space}%
\providecommand \EOS [0]{\spacefactor3000\relax}%
\providecommand \BibitemShut  [1]{\csname bibitem#1\endcsname}%
\let\auto@bib@innerbib\@empty
\bibitem [{\citenamefont {Giovannetti}\ \emph {et~al.}(2011)\citenamefont
  {Giovannetti}, \citenamefont {Lloyd},\ and\ \citenamefont
  {Maccone}}]{Giovannetti2011}%
  \BibitemOpen
  \bibfield  {author} {\bibinfo {author} {\bibfnamefont {V.}~\bibnamefont
  {Giovannetti}}, \bibinfo {author} {\bibfnamefont {S.}~\bibnamefont {Lloyd}},
  \ and\ \bibinfo {author} {\bibfnamefont {L.}~\bibnamefont {Maccone}},\ }\href
  {\doibase 10.1038/nphoton.2011.35} {\bibfield  {journal} {\bibinfo  {journal}
  {Nat. Photon.}\ }\textbf {\bibinfo {volume} {5}},\ \bibinfo {pages} {222}
  (\bibinfo {year} {2011})}\BibitemShut {NoStop}%
\bibitem [{\citenamefont {Holzwarth}\ \emph {et~al.}(2000)\citenamefont
  {Holzwarth}, \citenamefont {Udem}, \citenamefont {H{\"a}nsch}, \citenamefont
  {Knight}, \citenamefont {Wadsworth},\ and\ \citenamefont
  {Russell}}]{Holzwarth2000}%
  \BibitemOpen
  \bibfield  {author} {\bibinfo {author} {\bibfnamefont {R.}~\bibnamefont
  {Holzwarth}}, \bibinfo {author} {\bibfnamefont {T.}~\bibnamefont {Udem}},
  \bibinfo {author} {\bibfnamefont {T.~W.}\ \bibnamefont {H{\"a}nsch}},
  \bibinfo {author} {\bibfnamefont {J.~C.}\ \bibnamefont {Knight}}, \bibinfo
  {author} {\bibfnamefont {W.~J.}\ \bibnamefont {Wadsworth}}, \ and\ \bibinfo
  {author} {\bibfnamefont {P.~S.~J.}\ \bibnamefont {Russell}},\ }\href
  {\doibase 10.1103/PhysRevLett.85.2264} {\bibfield  {journal} {\bibinfo
  {journal} {Phys. Rev. Lett.}\ }\textbf {\bibinfo {volume} {85}},\ \bibinfo
  {pages} {2264} (\bibinfo {year} {2000})}\BibitemShut {NoStop}%
\bibitem [{\citenamefont {Del'Haye}\ \emph {et~al.}(2007)\citenamefont
  {Del'Haye}, \citenamefont {Schliesser}, \citenamefont {Arcizet},
  \citenamefont {Wilken}, \citenamefont {Holzwarth},\ and\ \citenamefont
  {Kippenberg}}]{DelHaye2007}%
  \BibitemOpen
  \bibfield  {author} {\bibinfo {author} {\bibfnamefont {P.}~\bibnamefont
  {Del'Haye}}, \bibinfo {author} {\bibfnamefont {A.}~\bibnamefont
  {Schliesser}}, \bibinfo {author} {\bibfnamefont {O.}~\bibnamefont {Arcizet}},
  \bibinfo {author} {\bibfnamefont {T.}~\bibnamefont {Wilken}}, \bibinfo
  {author} {\bibfnamefont {R.}~\bibnamefont {Holzwarth}}, \ and\ \bibinfo
  {author} {\bibfnamefont {T.~J.}\ \bibnamefont {Kippenberg}},\ }\href
  {\doibase 10.1038/nature06401} {\bibfield  {journal} {\bibinfo  {journal}
  {Nature}\ }\textbf {\bibinfo {volume} {450}},\ \bibinfo {pages} {1214}
  (\bibinfo {year} {2007})}\BibitemShut {NoStop}%
\bibitem [{\citenamefont {Pinel}\ \emph {et~al.}(2012)\citenamefont {Pinel},
  \citenamefont {Jian}, \citenamefont {de~Ara{\'u}jo}, \citenamefont {Feng},
  \citenamefont {Chalopin}, \citenamefont {Fabre},\ and\ \citenamefont
  {Treps}}]{Pinel2012}%
  \BibitemOpen
  \bibfield  {author} {\bibinfo {author} {\bibfnamefont {O.}~\bibnamefont
  {Pinel}}, \bibinfo {author} {\bibfnamefont {P.}~\bibnamefont {Jian}},
  \bibinfo {author} {\bibfnamefont {R.~M.}\ \bibnamefont {de~Ara{\'u}jo}},
  \bibinfo {author} {\bibfnamefont {J.}~\bibnamefont {Feng}}, \bibinfo {author}
  {\bibfnamefont {B.}~\bibnamefont {Chalopin}}, \bibinfo {author}
  {\bibfnamefont {C.}~\bibnamefont {Fabre}}, \ and\ \bibinfo {author}
  {\bibfnamefont {N.}~\bibnamefont {Treps}},\ }\href {\doibase
  10.1103/PhysRevLett.108.083601} {\bibfield  {journal} {\bibinfo  {journal}
  {Phys. Rev. Lett.}\ }\textbf {\bibinfo {volume} {108}},\ \bibinfo {pages}
  {083601} (\bibinfo {year} {2012})}\BibitemShut {NoStop}%
\bibitem [{\citenamefont {Lvovsky}\ \emph {et~al.}(2009)\citenamefont
  {Lvovsky}, \citenamefont {Sanders},\ and\ \citenamefont
  {Tittel}}]{Lvovsky2009}%
  \BibitemOpen
  \bibfield  {author} {\bibinfo {author} {\bibfnamefont {A.~I.}\ \bibnamefont
  {Lvovsky}}, \bibinfo {author} {\bibfnamefont {B.~C.}\ \bibnamefont
  {Sanders}}, \ and\ \bibinfo {author} {\bibfnamefont {W.}~\bibnamefont
  {Tittel}},\ }\href {\doibase 10.1038/nphoton.2009.231} {\bibfield  {journal}
  {\bibinfo  {journal} {Nat. Photon.}\ }\textbf {\bibinfo {volume} {3}},\
  \bibinfo {pages} {706} (\bibinfo {year} {2009})}\BibitemShut {NoStop}%
\bibitem [{\citenamefont {Reiserer}\ and\ \citenamefont
  {Rempe}(2015)}]{Reiserer2015}%
  \BibitemOpen
  \bibfield  {author} {\bibinfo {author} {\bibfnamefont {A.}~\bibnamefont
  {Reiserer}}\ and\ \bibinfo {author} {\bibfnamefont {G.}~\bibnamefont
  {Rempe}},\ }\href {\doibase 10.1103/RevModPhys.87.1379} {\bibfield  {journal}
  {\bibinfo  {journal} {Rev. Mod. Phys.}\ }\textbf {\bibinfo {volume} {87}},\
  \bibinfo {pages} {1379} (\bibinfo {year} {2015})}\BibitemShut {NoStop}%
\bibitem [{\citenamefont {Srivastava}\ \emph {et~al.}(2015)\citenamefont
  {Srivastava}, \citenamefont {Sidler}, \citenamefont {Allain}, \citenamefont
  {Lembke}, \citenamefont {Kis},\ and\ \citenamefont {Imamo{\u
  g}lu}}]{Srivastava2015}%
  \BibitemOpen
  \bibfield  {author} {\bibinfo {author} {\bibfnamefont {A.}~\bibnamefont
  {Srivastava}}, \bibinfo {author} {\bibfnamefont {M.}~\bibnamefont {Sidler}},
  \bibinfo {author} {\bibfnamefont {A.~V.}\ \bibnamefont {Allain}}, \bibinfo
  {author} {\bibfnamefont {D.~S.}\ \bibnamefont {Lembke}}, \bibinfo {author}
  {\bibfnamefont {A.}~\bibnamefont {Kis}}, \ and\ \bibinfo {author}
  {\bibfnamefont {A.}~\bibnamefont {Imamo{\u g}lu}},\ }\href {\doibase
  10.1038/nnano.2015.60} {\bibfield  {journal} {\bibinfo  {journal} {Nat.
  Nanotechnol.}\ }\textbf {\bibinfo {volume} {10}},\ \bibinfo {pages} {491}
  (\bibinfo {year} {2015})}\BibitemShut {NoStop}%
\bibitem [{\citenamefont {Tran}\ \emph {et~al.}(2016)\citenamefont {Tran},
  \citenamefont {Bray}, \citenamefont {Ford}, \citenamefont {Toth},\ and\
  \citenamefont {Aharonovich}}]{Tran2016a}%
  \BibitemOpen
  \bibfield  {author} {\bibinfo {author} {\bibfnamefont {T.~T.}\ \bibnamefont
  {Tran}}, \bibinfo {author} {\bibfnamefont {K.}~\bibnamefont {Bray}}, \bibinfo
  {author} {\bibfnamefont {M.~J.}\ \bibnamefont {Ford}}, \bibinfo {author}
  {\bibfnamefont {M.}~\bibnamefont {Toth}}, \ and\ \bibinfo {author}
  {\bibfnamefont {I.}~\bibnamefont {Aharonovich}},\ }\href {\doibase
  10.1038/nnano.2015.242} {\bibfield  {journal} {\bibinfo  {journal} {Nat.
  Nanotech.}\ }\textbf {\bibinfo {volume} {11}},\ \bibinfo {pages} {37}
  (\bibinfo {year} {2016})}\BibitemShut {NoStop}%
\bibitem [{\citenamefont {Abdi}\ \emph {et~al.}(2017)\citenamefont {Abdi},
  \citenamefont {Hwang}, \citenamefont {Aghtar},\ and\ \citenamefont
  {Plenio}}]{Abdi2017}%
  \BibitemOpen
  \bibfield  {author} {\bibinfo {author} {\bibfnamefont {M.}~\bibnamefont
  {Abdi}}, \bibinfo {author} {\bibfnamefont {M.-J.}\ \bibnamefont {Hwang}},
  \bibinfo {author} {\bibfnamefont {M.}~\bibnamefont {Aghtar}}, \ and\ \bibinfo
  {author} {\bibfnamefont {M.~B.}\ \bibnamefont {Plenio}},\ }\href {\doibase
  10.1103/PhysRevLett.119.233602} {\bibfield  {journal} {\bibinfo  {journal}
  {Phys. Rev. Lett.}\ }\textbf {\bibinfo {volume} {119}},\ \bibinfo {pages}
  {233602} (\bibinfo {year} {2017})}\BibitemShut {NoStop}%
\bibitem [{\citenamefont {Aharonovich}\ and\ \citenamefont
  {Toth}(2017)}]{Aharonovich2017}%
  \BibitemOpen
  \bibfield  {author} {\bibinfo {author} {\bibfnamefont {I.}~\bibnamefont
  {Aharonovich}}\ and\ \bibinfo {author} {\bibfnamefont {M.}~\bibnamefont
  {Toth}},\ }\href {\doibase 10.1126/science.aao6951} {\bibfield  {journal}
  {\bibinfo  {journal} {Science}\ }\textbf {\bibinfo {volume} {358}},\ \bibinfo
  {pages} {170} (\bibinfo {year} {2017})}\BibitemShut {NoStop}%
\bibitem [{\citenamefont {Wilson-Rae}\ \emph {et~al.}(2004)\citenamefont
  {Wilson-Rae}, \citenamefont {Zoller},\ and\ \citenamefont {Imamo{\u
  g}lu}}]{Wilson-Rae2004}%
  \BibitemOpen
  \bibfield  {author} {\bibinfo {author} {\bibfnamefont {I.}~\bibnamefont
  {Wilson-Rae}}, \bibinfo {author} {\bibfnamefont {P.}~\bibnamefont {Zoller}},
  \ and\ \bibinfo {author} {\bibfnamefont {A.}~\bibnamefont {Imamo{\u g}lu}},\
  }\href {\doibase 10.1103/PhysRevLett.92.075507} {\bibfield  {journal}
  {\bibinfo  {journal} {Phys. Rev. Lett.}\ }\textbf {\bibinfo {volume} {92}},\
  \bibinfo {pages} {075507} (\bibinfo {year} {2004})}\BibitemShut {NoStop}%
\bibitem [{\citenamefont {Kepesidis}\ \emph {et~al.}(2013)\citenamefont
  {Kepesidis}, \citenamefont {Bennett}, \citenamefont {Portolan}, \citenamefont
  {Lukin},\ and\ \citenamefont {Rabl}}]{Kepesidis2013}%
  \BibitemOpen
  \bibfield  {author} {\bibinfo {author} {\bibfnamefont {K.~V.}\ \bibnamefont
  {Kepesidis}}, \bibinfo {author} {\bibfnamefont {S.~D.}\ \bibnamefont
  {Bennett}}, \bibinfo {author} {\bibfnamefont {S.}~\bibnamefont {Portolan}},
  \bibinfo {author} {\bibfnamefont {M.~D.}\ \bibnamefont {Lukin}}, \ and\
  \bibinfo {author} {\bibfnamefont {P.}~\bibnamefont {Rabl}},\ }\href {\doibase
  10.1103/PhysRevB.88.064105} {\bibfield  {journal} {\bibinfo  {journal} {Phys.
  Rev. B}\ }\textbf {\bibinfo {volume} {88}},\ \bibinfo {pages} {064105}
  (\bibinfo {year} {2013})}\BibitemShut {NoStop}%
\bibitem [{\citenamefont {Ovartchaiyapong}\ \emph {et~al.}(2014)\citenamefont
  {Ovartchaiyapong}, \citenamefont {Lee}, \citenamefont {Myers},\ and\
  \citenamefont {Jayich}}]{Ovartchaiyapong2014}%
  \BibitemOpen
  \bibfield  {author} {\bibinfo {author} {\bibfnamefont {P.}~\bibnamefont
  {Ovartchaiyapong}}, \bibinfo {author} {\bibfnamefont {K.~W.}\ \bibnamefont
  {Lee}}, \bibinfo {author} {\bibfnamefont {B.~A.}\ \bibnamefont {Myers}}, \
  and\ \bibinfo {author} {\bibfnamefont {A.~C.~B.}\ \bibnamefont {Jayich}},\
  }\href {\doibase 10.1038/ncomms5429} {\bibfield  {journal} {\bibinfo
  {journal} {Nat. Commun.}\ }\textbf {\bibinfo {volume} {5}},\ \bibinfo {pages}
  {4429} (\bibinfo {year} {2014})}\BibitemShut {NoStop}%
\bibitem [{\citenamefont {Barfuss}\ \emph {et~al.}(2015)\citenamefont
  {Barfuss}, \citenamefont {Teissier}, \citenamefont {Neu}, \citenamefont
  {Nunnenkamp},\ and\ \citenamefont {Maletinsky}}]{Barfuss2015}%
  \BibitemOpen
  \bibfield  {author} {\bibinfo {author} {\bibfnamefont {A.}~\bibnamefont
  {Barfuss}}, \bibinfo {author} {\bibfnamefont {J.}~\bibnamefont {Teissier}},
  \bibinfo {author} {\bibfnamefont {E.}~\bibnamefont {Neu}}, \bibinfo {author}
  {\bibfnamefont {A.}~\bibnamefont {Nunnenkamp}}, \ and\ \bibinfo {author}
  {\bibfnamefont {P.}~\bibnamefont {Maletinsky}},\ }\href
  {http://dx.doi.org/10.1038/nphys3411} {\bibfield  {journal} {\bibinfo
  {journal} {Nat. Phys.}\ }\textbf {\bibinfo {volume} {11}},\ \bibinfo {pages}
  {820} (\bibinfo {year} {2015})}\BibitemShut {NoStop}%
\bibitem [{\citenamefont {Lemonde}\ \emph {et~al.}(2018)\citenamefont
  {Lemonde}, \citenamefont {Meesala}, \citenamefont {Sipahigil}, \citenamefont
  {Schuetz}, \citenamefont {Lukin}, \citenamefont {Loncar},\ and\ \citenamefont
  {Rabl}}]{Lemonde2018}%
  \BibitemOpen
  \bibfield  {author} {\bibinfo {author} {\bibfnamefont {M.-A.}\ \bibnamefont
  {Lemonde}}, \bibinfo {author} {\bibfnamefont {S.}~\bibnamefont {Meesala}},
  \bibinfo {author} {\bibfnamefont {A.}~\bibnamefont {Sipahigil}}, \bibinfo
  {author} {\bibfnamefont {M.~J.~A.}\ \bibnamefont {Schuetz}}, \bibinfo
  {author} {\bibfnamefont {M.~D.}\ \bibnamefont {Lukin}}, \bibinfo {author}
  {\bibfnamefont {M.}~\bibnamefont {Loncar}}, \ and\ \bibinfo {author}
  {\bibfnamefont {P.}~\bibnamefont {Rabl}},\ }\href {\doibase
  10.1103/PhysRevLett.120.213603} {\bibfield  {journal} {\bibinfo  {journal}
  {Phys. Rev. Lett.}\ }\textbf {\bibinfo {volume} {120}},\ \bibinfo {pages}
  {213603} (\bibinfo {year} {2018})}\BibitemShut {NoStop}%
\bibitem [{\citenamefont {Moser}\ \emph {et~al.}(2014)\citenamefont {Moser},
  \citenamefont {Eichler}, \citenamefont {G{\"u}ttinger}, \citenamefont
  {Dykman},\ and\ \citenamefont {Bachtold}}]{Moser2014}%
  \BibitemOpen
  \bibfield  {author} {\bibinfo {author} {\bibfnamefont {J.}~\bibnamefont
  {Moser}}, \bibinfo {author} {\bibfnamefont {A.}~\bibnamefont {Eichler}},
  \bibinfo {author} {\bibfnamefont {J.}~\bibnamefont {G{\"u}ttinger}}, \bibinfo
  {author} {\bibfnamefont {M.~I.}\ \bibnamefont {Dykman}}, \ and\ \bibinfo
  {author} {\bibfnamefont {A.}~\bibnamefont {Bachtold}},\ }\href {\doibase
  10.1038/nnano.2014.234} {\bibfield  {journal} {\bibinfo  {journal} {Nat.
  Nanotechnol.}\ }\textbf {\bibinfo {volume} {9}},\ \bibinfo {pages} {1007}
  (\bibinfo {year} {2014})}\BibitemShut {NoStop}%
\bibitem [{\citenamefont {Singh}\ \emph {et~al.}(2014)\citenamefont {Singh},
  \citenamefont {Bosman}, \citenamefont {Schneider}, \citenamefont {Blanter},
  \citenamefont {Castellanos-Gomez},\ and\ \citenamefont {Steele}}]{Singh2014}%
  \BibitemOpen
  \bibfield  {author} {\bibinfo {author} {\bibfnamefont {V.}~\bibnamefont
  {Singh}}, \bibinfo {author} {\bibfnamefont {S.~J.}\ \bibnamefont {Bosman}},
  \bibinfo {author} {\bibfnamefont {B.~H.}\ \bibnamefont {Schneider}}, \bibinfo
  {author} {\bibfnamefont {Y.~M.}\ \bibnamefont {Blanter}}, \bibinfo {author}
  {\bibfnamefont {A.}~\bibnamefont {Castellanos-Gomez}}, \ and\ \bibinfo
  {author} {\bibfnamefont {G.~A.}\ \bibnamefont {Steele}},\ }\href {\doibase
  10.1038/nnano.2014.168} {\bibfield  {journal} {\bibinfo  {journal} {Nat.
  Nanotechnol.}\ }\textbf {\bibinfo {volume} {9}},\ \bibinfo {pages} {820}
  (\bibinfo {year} {2014})}\BibitemShut {NoStop}%
\bibitem [{\citenamefont {Muschik}\ \emph {et~al.}(2014)\citenamefont
  {Muschik}, \citenamefont {Moulieras}, \citenamefont {Bachtold}, \citenamefont
  {Koppens}, \citenamefont {Lewenstein},\ and\ \citenamefont
  {Chang}}]{Muschik2014}%
  \BibitemOpen
  \bibfield  {author} {\bibinfo {author} {\bibfnamefont {C.~A.}\ \bibnamefont
  {Muschik}}, \bibinfo {author} {\bibfnamefont {S.}~\bibnamefont {Moulieras}},
  \bibinfo {author} {\bibfnamefont {A.}~\bibnamefont {Bachtold}}, \bibinfo
  {author} {\bibfnamefont {F.~H.~L.}\ \bibnamefont {Koppens}}, \bibinfo
  {author} {\bibfnamefont {M.}~\bibnamefont {Lewenstein}}, \ and\ \bibinfo
  {author} {\bibfnamefont {D.~E.}\ \bibnamefont {Chang}},\ }\href {\doibase
  10.1103/PhysRevLett.112.223601} {\bibfield  {journal} {\bibinfo  {journal}
  {Phys. Rev. Lett.}\ }\textbf {\bibinfo {volume} {112}},\ \bibinfo {pages}
  {223601} (\bibinfo {year} {2014})}\BibitemShut {NoStop}%
\bibitem [{\citenamefont {Castellanos-Gomez}\ \emph {et~al.}(2015)\citenamefont
  {Castellanos-Gomez}, \citenamefont {Singh}, \citenamefont {van~der Zant},\
  and\ \citenamefont {Steele}}]{Castellanos-Gomez2015}%
  \BibitemOpen
  \bibfield  {author} {\bibinfo {author} {\bibfnamefont {A.}~\bibnamefont
  {Castellanos-Gomez}}, \bibinfo {author} {\bibfnamefont {V.}~\bibnamefont
  {Singh}}, \bibinfo {author} {\bibfnamefont {H.~S.~J.}\ \bibnamefont {van~der
  Zant}}, \ and\ \bibinfo {author} {\bibfnamefont {G.~A.}\ \bibnamefont
  {Steele}},\ }\href {\doibase 10.1002/andp.201400153} {\bibfield  {journal}
  {\bibinfo  {journal} {Ann. Phys. (Berl.)}\ }\textbf {\bibinfo {volume}
  {527}},\ \bibinfo {pages} {27} (\bibinfo {year} {2015})}\BibitemShut
  {NoStop}%
\bibitem [{\citenamefont {Reserbat-Plantey}\ \emph {et~al.}(2016)\citenamefont
  {Reserbat-Plantey}, \citenamefont {Sch{\"a}dler}, \citenamefont {Gaudreau},
  \citenamefont {Navickaite}, \citenamefont {G{\"u}ttinger}, \citenamefont
  {Chang}, \citenamefont {Toninelli}, \citenamefont {Bachtold},\ and\
  \citenamefont {Koppens}}]{Reserbat2016}%
  \BibitemOpen
  \bibfield  {author} {\bibinfo {author} {\bibfnamefont {A.}~\bibnamefont
  {Reserbat-Plantey}}, \bibinfo {author} {\bibfnamefont {K.~G.}\ \bibnamefont
  {Sch{\"a}dler}}, \bibinfo {author} {\bibfnamefont {L.}~\bibnamefont
  {Gaudreau}}, \bibinfo {author} {\bibfnamefont {G.}~\bibnamefont
  {Navickaite}}, \bibinfo {author} {\bibfnamefont {J.}~\bibnamefont
  {G{\"u}ttinger}}, \bibinfo {author} {\bibfnamefont {D.}~\bibnamefont
  {Chang}}, \bibinfo {author} {\bibfnamefont {C.}~\bibnamefont {Toninelli}},
  \bibinfo {author} {\bibfnamefont {A.}~\bibnamefont {Bachtold}}, \ and\
  \bibinfo {author} {\bibfnamefont {F.~H.~L.}\ \bibnamefont {Koppens}},\ }\href
  {\doibase 10.1038/ncomms10218} {\bibfield  {journal} {\bibinfo  {journal}
  {Nat. Commun.}\ }\textbf {\bibinfo {volume} {7}},\ \bibinfo {pages} {10218}
  (\bibinfo {year} {2016})}\BibitemShut {NoStop}%
\bibitem [{\citenamefont {Morell}\ \emph {et~al.}(2016)\citenamefont {Morell},
  \citenamefont {Reserbat-Plantey}, \citenamefont {Tsioutsios}, \citenamefont
  {Sch{\"a}dler}, \citenamefont {Dubin}, \citenamefont {Koppens},\ and\
  \citenamefont {Bachtold}}]{Morell2016}%
  \BibitemOpen
  \bibfield  {author} {\bibinfo {author} {\bibfnamefont {N.}~\bibnamefont
  {Morell}}, \bibinfo {author} {\bibfnamefont {A.}~\bibnamefont
  {Reserbat-Plantey}}, \bibinfo {author} {\bibfnamefont {I.}~\bibnamefont
  {Tsioutsios}}, \bibinfo {author} {\bibfnamefont {K.~G.}\ \bibnamefont
  {Sch{\"a}dler}}, \bibinfo {author} {\bibfnamefont {F.}~\bibnamefont {Dubin}},
  \bibinfo {author} {\bibfnamefont {F.~H.~L.}\ \bibnamefont {Koppens}}, \ and\
  \bibinfo {author} {\bibfnamefont {A.}~\bibnamefont {Bachtold}},\ }\href
  {\doibase 10.1021/acs.nanolett.6b02038} {\bibfield  {journal} {\bibinfo
  {journal} {Nano Lett.}\ }\textbf {\bibinfo {volume} {16}},\ \bibinfo {pages}
  {5102} (\bibinfo {year} {2016})}\BibitemShut {NoStop}%
\bibitem [{\citenamefont {Li}\ and\ \citenamefont {Chen}(2016)}]{Li2016}%
  \BibitemOpen
  \bibfield  {author} {\bibinfo {author} {\bibfnamefont {L.~H.}\ \bibnamefont
  {Li}}\ and\ \bibinfo {author} {\bibfnamefont {Y.}~\bibnamefont {Chen}},\
  }\href {\doibase 10.1002/adfm.201504606} {\bibfield  {journal} {\bibinfo
  {journal} {Adv. Funct. Mater.}\ }\textbf {\bibinfo {volume} {26}},\ \bibinfo
  {pages} {2594} (\bibinfo {year} {2016})}\BibitemShut {NoStop}%
\bibitem [{\citenamefont {Falin}\ \emph {et~al.}(2017)\citenamefont {Falin},
  \citenamefont {Cai}, \citenamefont {Santos}, \citenamefont {Scullion},
  \citenamefont {Qian}, \citenamefont {Zhang}, \citenamefont {Yang},
  \citenamefont {Huang}, \citenamefont {Watanabe}, \citenamefont {Taniguchi},
  \citenamefont {Barnett}, \citenamefont {Chen}, \citenamefont {Ruoff},\ and\
  \citenamefont {Li}}]{Falin2017}%
  \BibitemOpen
  \bibfield  {author} {\bibinfo {author} {\bibfnamefont {A.}~\bibnamefont
  {Falin}}, \bibinfo {author} {\bibfnamefont {Q.}~\bibnamefont {Cai}}, \bibinfo
  {author} {\bibfnamefont {E.~J.~G.}\ \bibnamefont {Santos}}, \bibinfo {author}
  {\bibfnamefont {D.}~\bibnamefont {Scullion}}, \bibinfo {author}
  {\bibfnamefont {D.}~\bibnamefont {Qian}}, \bibinfo {author} {\bibfnamefont
  {R.}~\bibnamefont {Zhang}}, \bibinfo {author} {\bibfnamefont
  {Z.}~\bibnamefont {Yang}}, \bibinfo {author} {\bibfnamefont {S.}~\bibnamefont
  {Huang}}, \bibinfo {author} {\bibfnamefont {K.}~\bibnamefont {Watanabe}},
  \bibinfo {author} {\bibfnamefont {T.}~\bibnamefont {Taniguchi}}, \bibinfo
  {author} {\bibfnamefont {M.~R.}\ \bibnamefont {Barnett}}, \bibinfo {author}
  {\bibfnamefont {Y.}~\bibnamefont {Chen}}, \bibinfo {author} {\bibfnamefont
  {R.~S.}\ \bibnamefont {Ruoff}}, \ and\ \bibinfo {author} {\bibfnamefont
  {L.~H.}\ \bibnamefont {Li}},\ }\href {\doibase 10.1038/ncomms15815}
  {\bibfield  {journal} {\bibinfo  {journal} {Nat. Commun.}\ }\textbf {\bibinfo
  {volume} {8}},\ \bibinfo {pages} {15815} (\bibinfo {year}
  {2017})}\BibitemShut {NoStop}%
\bibitem [{\citenamefont {Zheng}\ \emph {et~al.}(2017)\citenamefont {Zheng},
  \citenamefont {Lee},\ and\ \citenamefont {Feng}}]{Zheng2017}%
  \BibitemOpen
  \bibfield  {author} {\bibinfo {author} {\bibfnamefont {X.-Q.}\ \bibnamefont
  {Zheng}}, \bibinfo {author} {\bibfnamefont {J.}~\bibnamefont {Lee}}, \ and\
  \bibinfo {author} {\bibfnamefont {P.~X.~L.}\ \bibnamefont {Feng}},\ }\href
  {\doibase 10.1038/micronano.2017.38} {\bibfield  {journal} {\bibinfo
  {journal} {Microsyst. Nanoeng.}\ }\textbf {\bibinfo {volume} {3}},\ \bibinfo
  {pages} {17038} (\bibinfo {year} {2017})}\BibitemShut {NoStop}%
\bibitem [{\citenamefont {Bassi}\ \emph {et~al.}(2013)\citenamefont {Bassi},
  \citenamefont {Lochan}, \citenamefont {Satin}, \citenamefont {Singh},\ and\
  \citenamefont {Ulbricht}}]{Bassi2013}%
  \BibitemOpen
  \bibfield  {author} {\bibinfo {author} {\bibfnamefont {A.}~\bibnamefont
  {Bassi}}, \bibinfo {author} {\bibfnamefont {K.}~\bibnamefont {Lochan}},
  \bibinfo {author} {\bibfnamefont {S.}~\bibnamefont {Satin}}, \bibinfo
  {author} {\bibfnamefont {T.~P.}\ \bibnamefont {Singh}}, \ and\ \bibinfo
  {author} {\bibfnamefont {H.}~\bibnamefont {Ulbricht}},\ }\href {\doibase
  10.1103/RevModPhys.85.471} {\bibfield  {journal} {\bibinfo  {journal} {Rev.
  Mod. Phys.}\ }\textbf {\bibinfo {volume} {85}},\ \bibinfo {pages} {471}
  (\bibinfo {year} {2013})}\BibitemShut {NoStop}%
\bibitem [{\citenamefont {Riedinger}\ \emph {et~al.}(2016)\citenamefont
  {Riedinger}, \citenamefont {Hong}, \citenamefont {Norte}, \citenamefont
  {Slater}, \citenamefont {Shang}, \citenamefont {Krause}, \citenamefont
  {Anant}, \citenamefont {Aspelmeyer},\ and\ \citenamefont
  {Gr{\"o}blacher}}]{Riedinger2016}%
  \BibitemOpen
  \bibfield  {author} {\bibinfo {author} {\bibfnamefont {R.}~\bibnamefont
  {Riedinger}}, \bibinfo {author} {\bibfnamefont {S.}~\bibnamefont {Hong}},
  \bibinfo {author} {\bibfnamefont {R.~A.}\ \bibnamefont {Norte}}, \bibinfo
  {author} {\bibfnamefont {J.~A.}\ \bibnamefont {Slater}}, \bibinfo {author}
  {\bibfnamefont {J.}~\bibnamefont {Shang}}, \bibinfo {author} {\bibfnamefont
  {A.~G.}\ \bibnamefont {Krause}}, \bibinfo {author} {\bibfnamefont
  {V.}~\bibnamefont {Anant}}, \bibinfo {author} {\bibfnamefont
  {M.}~\bibnamefont {Aspelmeyer}}, \ and\ \bibinfo {author} {\bibfnamefont
  {S.}~\bibnamefont {Gr{\"o}blacher}},\ }\href {\doibase 10.1038/nature16536}
  {\bibfield  {journal} {\bibinfo  {journal} {Nature}\ }\textbf {\bibinfo
  {volume} {530}},\ \bibinfo {pages} {313} (\bibinfo {year}
  {2016})}\BibitemShut {NoStop}%
\bibitem [{\citenamefont {Abdi}\ \emph {et~al.}(2016)\citenamefont {Abdi},
  \citenamefont {Degenfeld-Schonburg}, \citenamefont {Sameti}, \citenamefont
  {Navarrete-Benlloch},\ and\ \citenamefont {Hartmann}}]{Abdi2016}%
  \BibitemOpen
  \bibfield  {author} {\bibinfo {author} {\bibfnamefont {M.}~\bibnamefont
  {Abdi}}, \bibinfo {author} {\bibfnamefont {P.}~\bibnamefont
  {Degenfeld-Schonburg}}, \bibinfo {author} {\bibfnamefont {M.}~\bibnamefont
  {Sameti}}, \bibinfo {author} {\bibfnamefont {C.}~\bibnamefont
  {Navarrete-Benlloch}}, \ and\ \bibinfo {author} {\bibfnamefont {M.~J.}\
  \bibnamefont {Hartmann}},\ }\href {\doibase 10.1103/PhysRevLett.116.233604}
  {\bibfield  {journal} {\bibinfo  {journal} {Phys. Rev. Lett.}\ }\textbf
  {\bibinfo {volume} {116}},\ \bibinfo {pages} {233604} (\bibinfo {year}
  {2016})}\BibitemShut {NoStop}%
\bibitem [{\citenamefont {Weis}\ \emph {et~al.}(2010)\citenamefont {Weis},
  \citenamefont {Rivi{\`e}re}, \citenamefont {Del{\'e}glise}, \citenamefont
  {Gavartin}, \citenamefont {Arcizet}, \citenamefont {Schliesser},\ and\
  \citenamefont {Kippenberg}}]{Weis2010}%
  \BibitemOpen
  \bibfield  {author} {\bibinfo {author} {\bibfnamefont {S.}~\bibnamefont
  {Weis}}, \bibinfo {author} {\bibfnamefont {R.}~\bibnamefont {Rivi{\`e}re}},
  \bibinfo {author} {\bibfnamefont {S.}~\bibnamefont {Del{\'e}glise}}, \bibinfo
  {author} {\bibfnamefont {E.}~\bibnamefont {Gavartin}}, \bibinfo {author}
  {\bibfnamefont {O.}~\bibnamefont {Arcizet}}, \bibinfo {author} {\bibfnamefont
  {A.}~\bibnamefont {Schliesser}}, \ and\ \bibinfo {author} {\bibfnamefont
  {T.~J.}\ \bibnamefont {Kippenberg}},\ }\href {\doibase
  10.1126/science.1195596} {\bibfield  {journal} {\bibinfo  {journal}
  {Science}\ }\textbf {\bibinfo {volume} {330}},\ \bibinfo {pages} {1520}
  (\bibinfo {year} {2010})}\BibitemShut {NoStop}%
\bibitem [{\citenamefont {Safavi-Naeini}\ \emph {et~al.}(2011)\citenamefont
  {Safavi-Naeini}, \citenamefont {Mayer~Alegre}, \citenamefont {Chan},
  \citenamefont {Eichenfield}, \citenamefont {Winger}, \citenamefont {Lin},
  \citenamefont {Hill}, \citenamefont {Chang},\ and\ \citenamefont
  {Painter}}]{Safavi2011}%
  \BibitemOpen
  \bibfield  {author} {\bibinfo {author} {\bibfnamefont {A.~H.}\ \bibnamefont
  {Safavi-Naeini}}, \bibinfo {author} {\bibfnamefont {T.~P.}\ \bibnamefont
  {Mayer~Alegre}}, \bibinfo {author} {\bibfnamefont {J.}~\bibnamefont {Chan}},
  \bibinfo {author} {\bibfnamefont {M.}~\bibnamefont {Eichenfield}}, \bibinfo
  {author} {\bibfnamefont {M.}~\bibnamefont {Winger}}, \bibinfo {author}
  {\bibfnamefont {Q.}~\bibnamefont {Lin}}, \bibinfo {author} {\bibfnamefont
  {J.~T.}\ \bibnamefont {Hill}}, \bibinfo {author} {\bibfnamefont {D.~E.}\
  \bibnamefont {Chang}}, \ and\ \bibinfo {author} {\bibfnamefont
  {O.}~\bibnamefont {Painter}},\ }\href {\doibase 10.1038/nature09933}
  {\bibfield  {journal} {\bibinfo  {journal} {Nature}\ }\textbf {\bibinfo
  {volume} {472}},\ \bibinfo {pages} {69} (\bibinfo {year} {2011})}\BibitemShut
  {NoStop}%
\bibitem [{\citenamefont {M{\"u}cke}\ \emph {et~al.}(2010)\citenamefont
  {M{\"u}cke}, \citenamefont {Figueroa}, \citenamefont {Bochmann},
  \citenamefont {Hahn}, \citenamefont {Murr}, \citenamefont {Ritter},
  \citenamefont {Villas-Boas},\ and\ \citenamefont {Rempe}}]{Mucke2010}%
  \BibitemOpen
  \bibfield  {author} {\bibinfo {author} {\bibfnamefont {M.}~\bibnamefont
  {M{\"u}cke}}, \bibinfo {author} {\bibfnamefont {E.}~\bibnamefont {Figueroa}},
  \bibinfo {author} {\bibfnamefont {J.}~\bibnamefont {Bochmann}}, \bibinfo
  {author} {\bibfnamefont {C.}~\bibnamefont {Hahn}}, \bibinfo {author}
  {\bibfnamefont {K.}~\bibnamefont {Murr}}, \bibinfo {author} {\bibfnamefont
  {S.}~\bibnamefont {Ritter}}, \bibinfo {author} {\bibfnamefont {C.~J.}\
  \bibnamefont {Villas-Boas}}, \ and\ \bibinfo {author} {\bibfnamefont
  {G.}~\bibnamefont {Rempe}},\ }\href {\doibase 10.1038/nature09093} {\bibfield
   {journal} {\bibinfo  {journal} {Nature}\ }\textbf {\bibinfo {volume}
  {465}},\ \bibinfo {pages} {755} (\bibinfo {year} {2010})}\BibitemShut
  {NoStop}%
\bibitem [{\citenamefont {Kampschulte}\ \emph {et~al.}(2010)\citenamefont
  {Kampschulte}, \citenamefont {Alt}, \citenamefont {Brakhane}, \citenamefont
  {Eckstein}, \citenamefont {Reimann}, \citenamefont {Widera},\ and\
  \citenamefont {Meschede}}]{Kampschulte2010}%
  \BibitemOpen
  \bibfield  {author} {\bibinfo {author} {\bibfnamefont {T.}~\bibnamefont
  {Kampschulte}}, \bibinfo {author} {\bibfnamefont {W.}~\bibnamefont {Alt}},
  \bibinfo {author} {\bibfnamefont {S.}~\bibnamefont {Brakhane}}, \bibinfo
  {author} {\bibfnamefont {M.}~\bibnamefont {Eckstein}}, \bibinfo {author}
  {\bibfnamefont {R.}~\bibnamefont {Reimann}}, \bibinfo {author} {\bibfnamefont
  {A.}~\bibnamefont {Widera}}, \ and\ \bibinfo {author} {\bibfnamefont
  {D.}~\bibnamefont {Meschede}},\ }\href {\doibase
  10.1103/PhysRevLett.105.153603} {\bibfield  {journal} {\bibinfo  {journal}
  {Phys. Rev. Lett.}\ }\textbf {\bibinfo {volume} {105}},\ \bibinfo {pages}
  {153603} (\bibinfo {year} {2010})}\BibitemShut {NoStop}%
\bibitem [{\citenamefont {Slodicka}\ \emph {et~al.}(2010)\citenamefont
  {Slodicka}, \citenamefont {Hetet}, \citenamefont {Gerber}, \citenamefont
  {Hennrich},\ and\ \citenamefont {Blatt}}]{Slodicka2010}%
  \BibitemOpen
  \bibfield  {author} {\bibinfo {author} {\bibfnamefont {L.}~\bibnamefont
  {Slodicka}}, \bibinfo {author} {\bibfnamefont {G.}~\bibnamefont {Hetet}},
  \bibinfo {author} {\bibfnamefont {S.}~\bibnamefont {Gerber}}, \bibinfo
  {author} {\bibfnamefont {M.}~\bibnamefont {Hennrich}}, \ and\ \bibinfo
  {author} {\bibfnamefont {R.}~\bibnamefont {Blatt}},\ }\href {\doibase
  10.1103/PhysRevLett.105.153604} {\bibfield  {journal} {\bibinfo  {journal}
  {Phys. Rev. Lett.}\ }\textbf {\bibinfo {volume} {105}},\ \bibinfo {pages}
  {153604} (\bibinfo {year} {2010})}\BibitemShut {NoStop}%
\bibitem [{\citenamefont {Sch{\"o}n}\ \emph {et~al.}(2005)\citenamefont
  {Sch{\"o}n}, \citenamefont {Solano}, \citenamefont {Verstraete},
  \citenamefont {Cirac},\ and\ \citenamefont {Wolf}}]{Schon2005}%
  \BibitemOpen
  \bibfield  {author} {\bibinfo {author} {\bibfnamefont {C.}~\bibnamefont
  {Sch{\"o}n}}, \bibinfo {author} {\bibfnamefont {E.}~\bibnamefont {Solano}},
  \bibinfo {author} {\bibfnamefont {F.}~\bibnamefont {Verstraete}}, \bibinfo
  {author} {\bibfnamefont {J.~I.}\ \bibnamefont {Cirac}}, \ and\ \bibinfo
  {author} {\bibfnamefont {M.~M.}\ \bibnamefont {Wolf}},\ }\href {\doibase
  10.1103/PhysRevLett.95.110503} {\bibfield  {journal} {\bibinfo  {journal}
  {Phys. Rev. Lett.}\ }\textbf {\bibinfo {volume} {95}},\ \bibinfo {pages}
  {110503} (\bibinfo {year} {2005})}\BibitemShut {NoStop}%
\bibitem [{\citenamefont {Sch{\"o}n}\ \emph {et~al.}(2007)\citenamefont
  {Sch{\"o}n}, \citenamefont {Hammerer}, \citenamefont {Wolf}, \citenamefont
  {Cirac},\ and\ \citenamefont {Solano}}]{Schon2007}%
  \BibitemOpen
  \bibfield  {author} {\bibinfo {author} {\bibfnamefont {C.}~\bibnamefont
  {Sch{\"o}n}}, \bibinfo {author} {\bibfnamefont {K.}~\bibnamefont {Hammerer}},
  \bibinfo {author} {\bibfnamefont {M.~M.}\ \bibnamefont {Wolf}}, \bibinfo
  {author} {\bibfnamefont {J.~I.}\ \bibnamefont {Cirac}}, \ and\ \bibinfo
  {author} {\bibfnamefont {E.}~\bibnamefont {Solano}},\ }\href {\doibase
  10.1103/PhysRevA.75.032311} {\bibfield  {journal} {\bibinfo  {journal} {Phys.
  Rev. A}\ }\textbf {\bibinfo {volume} {75}},\ \bibinfo {pages} {032311}
  (\bibinfo {year} {2007})}\BibitemShut {NoStop}%
\bibitem [{\citenamefont {Leibfried}\ \emph {et~al.}(2004)\citenamefont
  {Leibfried}, \citenamefont {Barrett}, \citenamefont {Schaetz}, \citenamefont
  {Britton}, \citenamefont {Chiaverini}, \citenamefont {Itano}, \citenamefont
  {Jost}, \citenamefont {Langer},\ and\ \citenamefont
  {Wineland}}]{Leibfried2004}%
  \BibitemOpen
  \bibfield  {author} {\bibinfo {author} {\bibfnamefont {D.}~\bibnamefont
  {Leibfried}}, \bibinfo {author} {\bibfnamefont {M.~D.}\ \bibnamefont
  {Barrett}}, \bibinfo {author} {\bibfnamefont {T.}~\bibnamefont {Schaetz}},
  \bibinfo {author} {\bibfnamefont {J.}~\bibnamefont {Britton}}, \bibinfo
  {author} {\bibfnamefont {J.}~\bibnamefont {Chiaverini}}, \bibinfo {author}
  {\bibfnamefont {W.~M.}\ \bibnamefont {Itano}}, \bibinfo {author}
  {\bibfnamefont {J.~D.}\ \bibnamefont {Jost}}, \bibinfo {author}
  {\bibfnamefont {C.}~\bibnamefont {Langer}}, \ and\ \bibinfo {author}
  {\bibfnamefont {D.~J.}\ \bibnamefont {Wineland}},\ }\href {\doibase
  10.1126/science.1097576} {\bibfield  {journal} {\bibinfo  {journal}
  {Science}\ }\textbf {\bibinfo {volume} {304}},\ \bibinfo {pages} {1476}
  (\bibinfo {year} {2004})}\BibitemShut {NoStop}%
\bibitem [{\citenamefont {l.~O'brien}\ \emph {et~al.}(2009)\citenamefont
  {l.~O'brien}, \citenamefont {Furusawa},\ and\ \citenamefont {Vu{\v
  c}kovi{\'c}}}]{Obrien2009}%
  \BibitemOpen
  \bibfield  {author} {\bibinfo {author} {\bibfnamefont {J.}~\bibnamefont
  {l.~O'brien}}, \bibinfo {author} {\bibfnamefont {A.}~\bibnamefont
  {Furusawa}}, \ and\ \bibinfo {author} {\bibfnamefont {J.}~\bibnamefont {Vu{\v
  c}kovi{\'c}}},\ }\href {\doibase 10.1038/nphoton.2009.229} {\bibfield
  {journal} {\bibinfo  {journal} {Nat. Photon.}\ }\textbf {\bibinfo {volume}
  {3}},\ \bibinfo {pages} {687} (\bibinfo {year} {2009})}\BibitemShut {NoStop}%
\bibitem [{\citenamefont {Reimer}\ \emph {et~al.}(2016)\citenamefont {Reimer},
  \citenamefont {Kues}, \citenamefont {Roztocki}, \citenamefont {Wetzel},
  \citenamefont {Grazioso}, \citenamefont {Little}, \citenamefont {Chu},
  \citenamefont {Johnston}, \citenamefont {Bromberg}, \citenamefont {Caspani},
  \citenamefont {Moss},\ and\ \citenamefont {Morandotti}}]{Reimer2016}%
  \BibitemOpen
  \bibfield  {author} {\bibinfo {author} {\bibfnamefont {C.}~\bibnamefont
  {Reimer}}, \bibinfo {author} {\bibfnamefont {M.}~\bibnamefont {Kues}},
  \bibinfo {author} {\bibfnamefont {P.}~\bibnamefont {Roztocki}}, \bibinfo
  {author} {\bibfnamefont {B.}~\bibnamefont {Wetzel}}, \bibinfo {author}
  {\bibfnamefont {F.}~\bibnamefont {Grazioso}}, \bibinfo {author}
  {\bibfnamefont {B.~E.}\ \bibnamefont {Little}}, \bibinfo {author}
  {\bibfnamefont {S.~T.}\ \bibnamefont {Chu}}, \bibinfo {author} {\bibfnamefont
  {T.}~\bibnamefont {Johnston}}, \bibinfo {author} {\bibfnamefont
  {Y.}~\bibnamefont {Bromberg}}, \bibinfo {author} {\bibfnamefont
  {L.}~\bibnamefont {Caspani}}, \bibinfo {author} {\bibfnamefont {D.~J.}\
  \bibnamefont {Moss}}, \ and\ \bibinfo {author} {\bibfnamefont
  {R.}~\bibnamefont {Morandotti}},\ }\href {\doibase 10.1126/science.aad8532}
  {\bibfield  {journal} {\bibinfo  {journal} {Science}\ }\textbf {\bibinfo
  {volume} {351}},\ \bibinfo {pages} {1176} (\bibinfo {year}
  {2016})}\BibitemShut {NoStop}%
\bibitem [{\citenamefont {Morigi}\ \emph {et~al.}(2000)\citenamefont {Morigi},
  \citenamefont {Eschner},\ and\ \citenamefont {Keitel}}]{Morigi2000}%
  \BibitemOpen
  \bibfield  {author} {\bibinfo {author} {\bibfnamefont {G.}~\bibnamefont
  {Morigi}}, \bibinfo {author} {\bibfnamefont {J.}~\bibnamefont {Eschner}}, \
  and\ \bibinfo {author} {\bibfnamefont {C.~H.}\ \bibnamefont {Keitel}},\
  }\href {\doibase 10.1103/PhysRevLett.85.4458} {\bibfield  {journal} {\bibinfo
   {journal} {Phys. Rev. Lett.}\ }\textbf {\bibinfo {volume} {85}},\ \bibinfo
  {pages} {4458} (\bibinfo {year} {2000})}\BibitemShut {NoStop}%
\bibitem [{\citenamefont {Abdi}\ \emph {et~al.}(2018)\citenamefont {Abdi},
  \citenamefont {Chou}, \citenamefont {Gali},\ and\ \citenamefont
  {Plenio}}]{Abdi2018}%
  \BibitemOpen
  \bibfield  {author} {\bibinfo {author} {\bibfnamefont {M.}~\bibnamefont
  {Abdi}}, \bibinfo {author} {\bibfnamefont {J.-P.}\ \bibnamefont {Chou}},
  \bibinfo {author} {\bibfnamefont {A.}~\bibnamefont {Gali}}, \ and\ \bibinfo
  {author} {\bibfnamefont {M.~B.}\ \bibnamefont {Plenio}},\ }\href {\doibase
  acsphotonics.7b01442} {\bibfield  {journal} {\bibinfo  {journal} {ACS
  Photonics}\ }\textbf {\bibinfo {volume} {5}},\ \bibinfo {pages} {1967}
  (\bibinfo {year} {2018})}\BibitemShut {NoStop}%
\bibitem [{\citenamefont {Buhmann}\ \emph {et~al.}(2004)\citenamefont
  {Buhmann}, \citenamefont {Kn{\"o}ll}, \citenamefont {Welsch},\ and\
  \citenamefont {Dung}}]{Buhmann2004}%
  \BibitemOpen
  \bibfield  {author} {\bibinfo {author} {\bibfnamefont {S.~Y.}\ \bibnamefont
  {Buhmann}}, \bibinfo {author} {\bibfnamefont {L.}~\bibnamefont {Kn{\"o}ll}},
  \bibinfo {author} {\bibfnamefont {D.-G.}\ \bibnamefont {Welsch}}, \ and\
  \bibinfo {author} {\bibfnamefont {H.~T.}\ \bibnamefont {Dung}},\ }\href
  {\doibase 10.1103/PhysRevA.70.052117} {\bibfield  {journal} {\bibinfo
  {journal} {Phys. Rev. A}\ }\textbf {\bibinfo {volume} {70}},\ \bibinfo
  {pages} {052117} (\bibinfo {year} {2004})}\BibitemShut {NoStop}%
\bibitem [{sup()}]{suppinfo}%
  \BibitemOpen
  \href@noop {} {\bibinfo  {journal} {See the supplementary information for a
  derivation of Casimir coupling, elasticity of the strip, laser heating and
  spectral diffusion effects, and more details about EIT including the
  derivation of Eq.~(\ref{EIT})~\cite{Dietrich2018, Scully1997, Meenehan2014}}\
  }\BibitemShut {NoStop}%
\bibitem [{\citenamefont {Landau}\ and\ \citenamefont
  {Lifshitz}(1975)}]{Landau1975}%
  \BibitemOpen
\bibfield  {journal} {  }\bibfield  {author} {\bibinfo {author} {\bibfnamefont
  {L.~D.}\ \bibnamefont {Landau}}\ and\ \bibinfo {author} {\bibfnamefont
  {E.~M.}\ \bibnamefont {Lifshitz}},\ }\href@noop {} {\emph {\bibinfo {title}
  {Theory of Elasticity}}}\ (\bibinfo  {publisher} {Pergamon Press},\ \bibinfo
  {year} {1975})\BibitemShut {NoStop}%
\bibitem [{\citenamefont {Jungwirth}\ \emph {et~al.}(2016)\citenamefont
  {Jungwirth}, \citenamefont {Calderon}, \citenamefont {Ji}, \citenamefont
  {Spencer}, \citenamefont {Flatt{\'e}},\ and\ \citenamefont
  {Fuchs}}]{Jungwirth2016}%
  \BibitemOpen
  \bibfield  {author} {\bibinfo {author} {\bibfnamefont {N.~R.}\ \bibnamefont
  {Jungwirth}}, \bibinfo {author} {\bibfnamefont {B.}~\bibnamefont {Calderon}},
  \bibinfo {author} {\bibfnamefont {Y.}~\bibnamefont {Ji}}, \bibinfo {author}
  {\bibfnamefont {M.~G.}\ \bibnamefont {Spencer}}, \bibinfo {author}
  {\bibfnamefont {M.~E.}\ \bibnamefont {Flatt{\'e}}}, \ and\ \bibinfo {author}
  {\bibfnamefont {G.~D.}\ \bibnamefont {Fuchs}},\ }\href {\doibase
  10.1021/acs.nanolett.6b01987} {\bibfield  {journal} {\bibinfo  {journal}
  {Nano Lett.}\ }\textbf {\bibinfo {volume} {16}},\ \bibinfo {pages} {6052}
  (\bibinfo {year} {2016})}\BibitemShut {NoStop}%
\bibitem [{\citenamefont {Rabl}(2010)}]{Rabl2010}%
  \BibitemOpen
  \bibfield  {author} {\bibinfo {author} {\bibfnamefont {P.}~\bibnamefont
  {Rabl}},\ }\href {\doibase 10.1103/PhysRevB.82.165320} {\bibfield  {journal}
  {\bibinfo  {journal} {Phys. Rev. B}\ }\textbf {\bibinfo {volume} {82}},\
  \bibinfo {pages} {165320} (\bibinfo {year} {2010})}\BibitemShut {NoStop}%
\bibitem [{\citenamefont {Degenfeld-Schonburg}\ \emph
  {et~al.}(2016)\citenamefont {Degenfeld-Schonburg}, \citenamefont {Abdi},
  \citenamefont {Hartmann},\ and\ \citenamefont
  {Navarrete-Benlloch}}]{Degenfeld2016}%
  \BibitemOpen
  \bibfield  {author} {\bibinfo {author} {\bibfnamefont {P.}~\bibnamefont
  {Degenfeld-Schonburg}}, \bibinfo {author} {\bibfnamefont {M.}~\bibnamefont
  {Abdi}}, \bibinfo {author} {\bibfnamefont {M.~J.}\ \bibnamefont {Hartmann}},
  \ and\ \bibinfo {author} {\bibfnamefont {C.}~\bibnamefont
  {Navarrete-Benlloch}},\ }\href {\doibase PhysRevA.93.023819} {\bibfield
  {journal} {\bibinfo  {journal} {Phys. Rev. A}\ }\textbf {\bibinfo {volume}
  {93}},\ \bibinfo {pages} {023819} (\bibinfo {year} {2016})}\BibitemShut
  {NoStop}%
\bibitem [{\citenamefont {Rebi{\'c}}\ \emph {et~al.}(2004)\citenamefont
  {Rebi{\'c}}, \citenamefont {Vitali}, \citenamefont {Ottaviani}, \citenamefont
  {Tombesi}, \citenamefont {Artoni}, \citenamefont {Cataliotti},\ and\
  \citenamefont {Corbal{\'a}n}}]{Rebic2004}%
  \BibitemOpen
  \bibfield  {author} {\bibinfo {author} {\bibfnamefont {S.}~\bibnamefont
  {Rebi{\'c}}}, \bibinfo {author} {\bibfnamefont {D.}~\bibnamefont {Vitali}},
  \bibinfo {author} {\bibfnamefont {C.}~\bibnamefont {Ottaviani}}, \bibinfo
  {author} {\bibfnamefont {P.}~\bibnamefont {Tombesi}}, \bibinfo {author}
  {\bibfnamefont {M.}~\bibnamefont {Artoni}}, \bibinfo {author} {\bibfnamefont
  {F.}~\bibnamefont {Cataliotti}}, \ and\ \bibinfo {author} {\bibfnamefont
  {R.}~\bibnamefont {Corbal{\'a}n}},\ }\href {\doibase
  10.1103/PhysRevA.70.032317} {\bibfield  {journal} {\bibinfo  {journal} {Phys.
  Rev. A}\ }\textbf {\bibinfo {volume} {70}},\ \bibinfo {pages} {032317}
  (\bibinfo {year} {2004})}\BibitemShut {NoStop}%
\bibitem [{\citenamefont {Cartamil-Bueno}\ \emph {et~al.}(2017)\citenamefont
  {Cartamil-Bueno}, \citenamefont {Cavalieri}, \citenamefont {Wang},
  \citenamefont {Houri}, \citenamefont {Hofmann},\ and\ \citenamefont {van~der
  Zant}}]{Cartamil-Bueno2017}%
  \BibitemOpen
  \bibfield  {author} {\bibinfo {author} {\bibfnamefont {S.~J.}\ \bibnamefont
  {Cartamil-Bueno}}, \bibinfo {author} {\bibfnamefont {M.}~\bibnamefont
  {Cavalieri}}, \bibinfo {author} {\bibfnamefont {R.}~\bibnamefont {Wang}},
  \bibinfo {author} {\bibfnamefont {S.}~\bibnamefont {Houri}}, \bibinfo
  {author} {\bibfnamefont {S.}~\bibnamefont {Hofmann}}, \ and\ \bibinfo
  {author} {\bibfnamefont {H.~S.~J.}\ \bibnamefont {van~der Zant}},\ }\href
  {\doibase 10.1038/s41699-017-0020-8} {\bibfield  {journal} {\bibinfo
  {journal} {npj 2D Mater. Appl.}\ }\textbf {\bibinfo {volume} {1}},\ \bibinfo
  {pages} {16} (\bibinfo {year} {2017})}\BibitemShut {NoStop}%
\bibitem [{\citenamefont {Shandilya}\ \emph {et~al.}(2018)\citenamefont
  {Shandilya}, \citenamefont {Fr{\"o}ch}, \citenamefont {Mitchell},
  \citenamefont {Lake}, \citenamefont {Kim}, \citenamefont {Toth},
  \citenamefont {Hajisalem}, \citenamefont {Aharonovich},\ and\ \citenamefont
  {Barclay}}]{Shandilya2018}%
  \BibitemOpen
  \bibfield  {author} {\bibinfo {author} {\bibfnamefont {P.~K.}\ \bibnamefont
  {Shandilya}}, \bibinfo {author} {\bibfnamefont {J.~E.}\ \bibnamefont
  {Fr{\"o}ch}}, \bibinfo {author} {\bibfnamefont {M.}~\bibnamefont {Mitchell}},
  \bibinfo {author} {\bibfnamefont {D.~P.}\ \bibnamefont {Lake}}, \bibinfo
  {author} {\bibfnamefont {S.}~\bibnamefont {Kim}}, \bibinfo {author}
  {\bibfnamefont {M.}~\bibnamefont {Toth}}, \bibinfo {author} {\bibfnamefont
  {G.}~\bibnamefont {Hajisalem}}, \bibinfo {author} {\bibfnamefont
  {I.}~\bibnamefont {Aharonovich}}, \ and\ \bibinfo {author} {\bibfnamefont
  {P.~E.}\ \bibnamefont {Barclay}},\ }\href {arXiv:1809.04023 [physics.optics]}
  {\bibfield  {journal} {\bibinfo  {journal} {arXiv:1809.04023
  [physics.optics]}\ } (\bibinfo {year} {2018})}\BibitemShut {NoStop}%
\bibitem [{\citenamefont {Meenehan}\ \emph {et~al.}(2015)\citenamefont
  {Meenehan}, \citenamefont {Cohen}, \citenamefont {MacCabe}, \citenamefont
  {Marsili}, \citenamefont {Shaw},\ and\ \citenamefont
  {Painter}}]{Meenehan2015}%
  \BibitemOpen
  \bibfield  {author} {\bibinfo {author} {\bibfnamefont {S.~M.}\ \bibnamefont
  {Meenehan}}, \bibinfo {author} {\bibfnamefont {J.~D.}\ \bibnamefont {Cohen}},
  \bibinfo {author} {\bibfnamefont {G.~S.}\ \bibnamefont {MacCabe}}, \bibinfo
  {author} {\bibfnamefont {F.}~\bibnamefont {Marsili}}, \bibinfo {author}
  {\bibfnamefont {M.~D.}\ \bibnamefont {Shaw}}, \ and\ \bibinfo {author}
  {\bibfnamefont {O.}~\bibnamefont {Painter}},\ }\href {\doibase
  10.1103/PhysRevX.5.041002} {\bibfield  {journal} {\bibinfo  {journal} {Phys.
  Rev. X}\ }\textbf {\bibinfo {volume} {5}},\ \bibinfo {pages} {041002}
  (\bibinfo {year} {2015})}\BibitemShut {NoStop}%
\bibitem [{\citenamefont {Wang}\ \emph {et~al.}(2014)\citenamefont {Wang},
  \citenamefont {Gu}, \citenamefont {Liu}, \citenamefont {Miranowicz},\ and\
  \citenamefont {Nori}}]{Wang2014}%
  \BibitemOpen
  \bibfield  {author} {\bibinfo {author} {\bibfnamefont {H.}~\bibnamefont
  {Wang}}, \bibinfo {author} {\bibfnamefont {X.}~\bibnamefont {Gu}}, \bibinfo
  {author} {\bibfnamefont {Y.}~\bibnamefont {Liu}}, \bibinfo {author}
  {\bibfnamefont {A.}~\bibnamefont {Miranowicz}}, \ and\ \bibinfo {author}
  {\bibfnamefont {F.}~\bibnamefont {Nori}},\ }\href {\doibase
  10.1103/PhysRevA.90.023817} {\bibfield  {journal} {\bibinfo  {journal} {Phys.
  Rev. A}\ }\textbf {\bibinfo {volume} {90}},\ \bibinfo {pages} {023817}
  (\bibinfo {year} {2014})}\BibitemShut {NoStop}%
\bibitem [{\citenamefont {Alotaibi}\ and\ \citenamefont
  {Sanders}(2014)}]{Alotaibi2014}%
  \BibitemOpen
  \bibfield  {author} {\bibinfo {author} {\bibfnamefont {H.~M.~M.}\
  \bibnamefont {Alotaibi}}\ and\ \bibinfo {author} {\bibfnamefont {B.~C.}\
  \bibnamefont {Sanders}},\ }\href {\doibase 10.1103/PhysRevA.89.021802}
  {\bibfield  {journal} {\bibinfo  {journal} {Phys. Rev. A}\ }\textbf {\bibinfo
  {volume} {89}},\ \bibinfo {pages} {021802} (\bibinfo {year}
  {2014})}\BibitemShut {NoStop}%
\bibitem [{\citenamefont {Kim}\ \emph {et~al.}(2018)\citenamefont {Kim},
  \citenamefont {Fr{\"o}ch}, \citenamefont {Christian}, \citenamefont {Straw},
  \citenamefont {Bishop}, \citenamefont {Totonjian}, \citenamefont {Watanabe},
  \citenamefont {Taniguchi}, \citenamefont {Toth},\ and\ \citenamefont
  {Aharonovich}}]{Kim2018}%
  \BibitemOpen
  \bibfield  {author} {\bibinfo {author} {\bibfnamefont {S.}~\bibnamefont
  {Kim}}, \bibinfo {author} {\bibfnamefont {J.~E.}\ \bibnamefont {Fr{\"o}ch}},
  \bibinfo {author} {\bibfnamefont {J.}~\bibnamefont {Christian}}, \bibinfo
  {author} {\bibfnamefont {M.}~\bibnamefont {Straw}}, \bibinfo {author}
  {\bibfnamefont {J.}~\bibnamefont {Bishop}}, \bibinfo {author} {\bibfnamefont
  {D.}~\bibnamefont {Totonjian}}, \bibinfo {author} {\bibfnamefont
  {K.}~\bibnamefont {Watanabe}}, \bibinfo {author} {\bibfnamefont
  {T.}~\bibnamefont {Taniguchi}}, \bibinfo {author} {\bibfnamefont
  {M.}~\bibnamefont {Toth}}, \ and\ \bibinfo {author} {\bibfnamefont
  {I.}~\bibnamefont {Aharonovich}},\ }\href {\doibase
  10.1038/s41467-018-05117-4} {\bibfield  {journal} {\bibinfo  {journal} {Nat.
  Commun.}\ }\textbf {\bibinfo {volume} {9}},\ \bibinfo {pages} {2623}
  (\bibinfo {year} {2018})}\BibitemShut {NoStop}%
\bibitem [{\citenamefont {Narducci}\ \emph {et~al.}(1990)\citenamefont
  {Narducci}, \citenamefont {Scully}, \citenamefont {Oppo}, \citenamefont
  {Ru},\ and\ \citenamefont {Tredicce}}]{Narducci1990}%
  \BibitemOpen
  \bibfield  {author} {\bibinfo {author} {\bibfnamefont {L.~M.}\ \bibnamefont
  {Narducci}}, \bibinfo {author} {\bibfnamefont {M.~O.}\ \bibnamefont
  {Scully}}, \bibinfo {author} {\bibfnamefont {G.-L.}\ \bibnamefont {Oppo}},
  \bibinfo {author} {\bibfnamefont {P.}~\bibnamefont {Ru}}, \ and\ \bibinfo
  {author} {\bibfnamefont {J.~R.}\ \bibnamefont {Tredicce}},\ }\href {\doibase
  10.1103/PhysRevA.42.1630} {\bibfield  {journal} {\bibinfo  {journal} {Phys.
  Rev. A}\ }\textbf {\bibinfo {volume} {42}},\ \bibinfo {pages} {1630}
  (\bibinfo {year} {1990})}\BibitemShut {NoStop}%
\bibitem [{\citenamefont {Carmichael}(1999)}]{Carmichael1999}%
  \BibitemOpen
  \bibfield  {author} {\bibinfo {author} {\bibfnamefont {H.~J.}\ \bibnamefont
  {Carmichael}},\ }\href@noop {} {\emph {\bibinfo {title} {Statistical Methods
  in Quantum Optics}}},\ Vol.~\bibinfo {volume} {1}\ (\bibinfo  {publisher}
  {Springer-Verlag},\ \bibinfo {address} {Heidelberg},\ \bibinfo {year}
  {1999})\BibitemShut {NoStop}%
\bibitem [{\citenamefont {Afzelius}\ and\ \citenamefont
  {Simon}(2010)}]{Afzelius2010}%
  \BibitemOpen
  \bibfield  {author} {\bibinfo {author} {\bibfnamefont {M.}~\bibnamefont
  {Afzelius}}\ and\ \bibinfo {author} {\bibfnamefont {C.}~\bibnamefont
  {Simon}},\ }\href {\doibase 10.1103/PhysRevA.82.022310} {\bibfield  {journal}
  {\bibinfo  {journal} {Phys. Rev. A}\ }\textbf {\bibinfo {volume} {82}},\
  \bibinfo {pages} {022310} (\bibinfo {year} {2010})}\BibitemShut {NoStop}%
\bibitem [{\citenamefont {Moiseev}\ \emph {et~al.}(2010)\citenamefont
  {Moiseev}, \citenamefont {Andrianov},\ and\ \citenamefont
  {Gubaidullin}}]{Moiseev2010}%
  \BibitemOpen
  \bibfield  {author} {\bibinfo {author} {\bibfnamefont {S.~A.}\ \bibnamefont
  {Moiseev}}, \bibinfo {author} {\bibfnamefont {S.~N.}\ \bibnamefont
  {Andrianov}}, \ and\ \bibinfo {author} {\bibfnamefont {F.~F.}\ \bibnamefont
  {Gubaidullin}},\ }\href {\doibase 10.1103/PhysRevA.82.022311} {\bibfield
  {journal} {\bibinfo  {journal} {Phys. Rev. A}\ }\textbf {\bibinfo {volume}
  {82}},\ \bibinfo {pages} {022311} (\bibinfo {year} {2010})}\BibitemShut
  {NoStop}%
\bibitem [{\citenamefont {Sabooni}\ \emph {et~al.}(2013)\citenamefont
  {Sabooni}, \citenamefont {Li}, \citenamefont {Kr{\"o}ll},\ and\ \citenamefont
  {Rippe}}]{Sabooni2013}%
  \BibitemOpen
  \bibfield  {author} {\bibinfo {author} {\bibfnamefont {M.}~\bibnamefont
  {Sabooni}}, \bibinfo {author} {\bibfnamefont {Q.}~\bibnamefont {Li}},
  \bibinfo {author} {\bibfnamefont {S.}~\bibnamefont {Kr{\"o}ll}}, \ and\
  \bibinfo {author} {\bibfnamefont {L.}~\bibnamefont {Rippe}},\ }\href
  {\doibase 10.1103/PhysRevLett.110.133604} {\bibfield  {journal} {\bibinfo
  {journal} {Phys. Rev. Lett.}\ }\textbf {\bibinfo {volume} {110}},\ \bibinfo
  {pages} {133604} (\bibinfo {year} {2013})}\BibitemShut {NoStop}%
\bibitem [{\citenamefont {Moiseev}\ and\ \citenamefont {{Le
  Gou{\"e}t}}(2012)}]{Moiseev2012}%
  \BibitemOpen
  \bibfield  {author} {\bibinfo {author} {\bibfnamefont {S.~A.}\ \bibnamefont
  {Moiseev}}\ and\ \bibinfo {author} {\bibfnamefont {J.-L.}\ \bibnamefont {{Le
  Gou{\"e}t}}},\ }\href {\doibase 10.1088/0953-4075/45/12/124003} {\bibfield
  {journal} {\bibinfo  {journal} {J. Phys. B: At. Mol. Opt. Phys.}\ }\textbf
  {\bibinfo {volume} {45}},\ \bibinfo {pages} {124003} (\bibinfo {year}
  {2012})}\BibitemShut {NoStop}%
\bibitem [{\citenamefont {Arsalanov}\ and\ \citenamefont
  {Moiseev}(2017)}]{Arsalanov2017}%
  \BibitemOpen
  \bibfield  {author} {\bibinfo {author} {\bibfnamefont {N.~M.}\ \bibnamefont
  {Arsalanov}}\ and\ \bibinfo {author} {\bibfnamefont {S.~A.}\ \bibnamefont
  {Moiseev}},\ }\href {\doibase /10.1070/QEL16467} {\bibfield  {journal}
  {\bibinfo  {journal} {Quantum Electron.}\ }\textbf {\bibinfo {volume} {47}},\
  \bibinfo {pages} {783} (\bibinfo {year} {2017})}\BibitemShut {NoStop}%
\bibitem [{\citenamefont {Rao}\ \emph {et~al.}(2015)\citenamefont {Rao},
  \citenamefont {Yang},\ and\ \citenamefont {Wrachtrup}}]{Rao2015}%
  \BibitemOpen
  \bibfield  {author} {\bibinfo {author} {\bibfnamefont {D.~D.~B.}\
  \bibnamefont {Rao}}, \bibinfo {author} {\bibfnamefont {S.}~\bibnamefont
  {Yang}}, \ and\ \bibinfo {author} {\bibfnamefont {J.}~\bibnamefont
  {Wrachtrup}},\ }\href {\doibase 10.1103/PhysRevB.92.081301} {\bibfield
  {journal} {\bibinfo  {journal} {Phys. Rev. B}\ }\textbf {\bibinfo {volume}
  {92}},\ \bibinfo {pages} {081301(R)} (\bibinfo {year} {2015})}\BibitemShut
  {NoStop}%
\bibitem [{\citenamefont {Dhand}\ \emph {et~al.}(2018)\citenamefont {Dhand},
  \citenamefont {Engelkemeier}, \citenamefont {Sansoni}, \citenamefont
  {Barkhofen}, \citenamefont {Silberhorn},\ and\ \citenamefont
  {Plenio}}]{Dhand2018}%
  \BibitemOpen
  \bibfield  {author} {\bibinfo {author} {\bibfnamefont {I.}~\bibnamefont
  {Dhand}}, \bibinfo {author} {\bibfnamefont {M.}~\bibnamefont {Engelkemeier}},
  \bibinfo {author} {\bibfnamefont {L.}~\bibnamefont {Sansoni}}, \bibinfo
  {author} {\bibfnamefont {S.}~\bibnamefont {Barkhofen}}, \bibinfo {author}
  {\bibfnamefont {C.}~\bibnamefont {Silberhorn}}, \ and\ \bibinfo {author}
  {\bibfnamefont {M.~B.}\ \bibnamefont {Plenio}},\ }\href {\doibase
  10.1103/PhysRevLett.120.130501} {\bibfield  {journal} {\bibinfo  {journal}
  {Phys. Rev. Lett.}\ }\textbf {\bibinfo {volume} {120}},\ \bibinfo {pages}
  {130501} (\bibinfo {year} {2018})}\BibitemShut {NoStop}%
\bibitem [{\citenamefont {Chen}\ \emph {et~al.}(2014)\citenamefont {Chen},
  \citenamefont {Menicucci},\ and\ \citenamefont {Pfister}}]{Chen2014}%
  \BibitemOpen
  \bibfield  {author} {\bibinfo {author} {\bibfnamefont {M.}~\bibnamefont
  {Chen}}, \bibinfo {author} {\bibfnamefont {N.~C.}\ \bibnamefont {Menicucci}},
  \ and\ \bibinfo {author} {\bibfnamefont {O.}~\bibnamefont {Pfister}},\ }\href
  {\doibase 10.1103/PhysRevLett.112.120505} {\bibfield  {journal} {\bibinfo
  {journal} {Phys. Rev. Lett.}\ }\textbf {\bibinfo {volume} {112}},\ \bibinfo
  {pages} {120505} (\bibinfo {year} {2014})}\BibitemShut {NoStop}%
\bibitem [{\citenamefont {Dietrich}\ \emph {et~al.}(2018)\citenamefont
  {Dietrich}, \citenamefont {B{\"u}rk}, \citenamefont {Steiger}, \citenamefont
  {Antoniuk}, \citenamefont {Tran}, \citenamefont {Nguyen}, \citenamefont
  {Aharonovich}, \citenamefont {Jelezko},\ and\ \citenamefont
  {Kubanek}}]{Dietrich2018}%
  \BibitemOpen
  \bibfield  {author} {\bibinfo {author} {\bibfnamefont {A.}~\bibnamefont
  {Dietrich}}, \bibinfo {author} {\bibfnamefont {M.}~\bibnamefont {B{\"u}rk}},
  \bibinfo {author} {\bibfnamefont {E.~S.}\ \bibnamefont {Steiger}}, \bibinfo
  {author} {\bibfnamefont {L.}~\bibnamefont {Antoniuk}}, \bibinfo {author}
  {\bibfnamefont {T.~T.}\ \bibnamefont {Tran}}, \bibinfo {author}
  {\bibfnamefont {M.}~\bibnamefont {Nguyen}}, \bibinfo {author} {\bibfnamefont
  {I.}~\bibnamefont {Aharonovich}}, \bibinfo {author} {\bibfnamefont
  {F.}~\bibnamefont {Jelezko}}, \ and\ \bibinfo {author} {\bibfnamefont
  {A.}~\bibnamefont {Kubanek}},\ }\href {\doibase 10.1103/PhysRevB.98.081414}
  {\bibfield  {journal} {\bibinfo  {journal} {Phys. Rev. B}\ }\textbf {\bibinfo
  {volume} {98}},\ \bibinfo {pages} {081414} (\bibinfo {year}
  {2018})}\BibitemShut {NoStop}%
\bibitem [{\citenamefont {Scully}\ and\ \citenamefont
  {Zubairy}(1997)}]{Scully1997}%
  \BibitemOpen
  \bibfield  {author} {\bibinfo {author} {\bibfnamefont {M.~O.}\ \bibnamefont
  {Scully}}\ and\ \bibinfo {author} {\bibfnamefont {M.~S.}\ \bibnamefont
  {Zubairy}},\ }\href@noop {} {\emph {\bibinfo {title} {Quantum Optics}}}\
  (\bibinfo  {publisher} {Cambridge University Press},\ \bibinfo {year}
  {1997})\BibitemShut {NoStop}%
\bibitem [{\citenamefont {Meenehan}\ \emph {et~al.}(2014)\citenamefont
  {Meenehan}, \citenamefont {Cohen}, \citenamefont {Gr{\"o}blacher},
  \citenamefont {Hill}, \citenamefont {Safavi-Naeini}, \citenamefont
  {Aspelmeyer},\ and\ \citenamefont {Painter}}]{Meenehan2014}%
  \BibitemOpen
  \bibfield  {author} {\bibinfo {author} {\bibfnamefont {S.~M.}\ \bibnamefont
  {Meenehan}}, \bibinfo {author} {\bibfnamefont {J.~D.}\ \bibnamefont {Cohen}},
  \bibinfo {author} {\bibfnamefont {S.}~\bibnamefont {Gr{\"o}blacher}},
  \bibinfo {author} {\bibfnamefont {J.~T.}\ \bibnamefont {Hill}}, \bibinfo
  {author} {\bibfnamefont {A.~H.}\ \bibnamefont {Safavi-Naeini}}, \bibinfo
  {author} {\bibfnamefont {M.}~\bibnamefont {Aspelmeyer}}, \ and\ \bibinfo
  {author} {\bibfnamefont {O.}~\bibnamefont {Painter}},\ }\href {\doibase
  10.1103/PhysRevA.90.011803} {\bibfield  {journal} {\bibinfo  {journal} {Phys.
  Rev. A}\ }\textbf {\bibinfo {volume} {90}},\ \bibinfo {pages} {011803(R)}
  (\bibinfo {year} {2014})}\BibitemShut {NoStop}%
\end{thebibliography}%

\end{document}